\documentclass[draft]{agujournal2019}
\usepackage{url} 
\usepackage{lineno}

\usepackage{soul}
\usepackage[finalnew]{trackchanges} 

\usepackage{amsmath}
\usepackage{amssymb}
\usepackage{booktabs}
\draftfalse

\journalname{JGR: Planets}

\begin{document}

%
%

\title{The efficient delivery of highly-siderophile elements to the core creates a mass accretion catastrophe for the Earth}

%
%

\authors{Richard J. Anslow\affil{1}, Maylis Landeau\affil{2}, Amy Bonsor\affil{1}, Jonathan Itcovitz\affil{1,3}, Oliver Shorttle\affil{1,4}}

\affiliation{1}{Institute of Astronomy, University of Cambridge, Madingley Road, Cambridge, CB3 0HA, UK}
\affiliation{2}{Université de Paris, Institut de Physique du Globe de Paris, CNRS, Paris, France}
\affiliation{3}{Department of Civil and Environmental Engineering, Imperial College London, London, SW7 2AZ, UK}
\affiliation{4}{Department of Earth Sciences, University of Cambridge, Downing Street, Cambridge, CB3 9ET, UK}

\correspondingauthor{Richard J. Anslow}{rja92@ast.cam.ac.uk}



\begin{keypoints}
\item The entrainment of metal, and its HSE content, in a magma pond is possible only for $\leq\,0.01\,$mm droplets.
\item Metal delivered by $\geq\,1\,$km impactors is lost to Earth’s core, yet constraints on total mass accretion prevents HSE delivery by $\leq\,1\,$km impactors.
\item Thus, either an oxidized late veneer or the disruption of large impactors into $\leq\,0.01\,$mm droplets is required to account for observed HSEs.
\end{keypoints}

%
%

%
%


\begin{abstract}
The excess abundance of highly siderophile elements (HSEs), as inferred for the terrestrial planets and the Moon, is thought to record a `late veneer' of impacts after the giant impact phase of planet formation. Estimates for total mass accretion during this period typically assume all HSEs delivered remain entrained in the mantle. Here, we present an analytical discussion of the fate of liquid metal diapirs in both a magma pond and a solid mantle, and show that metals from impactors larger than approximately 1\,km will sink to Earth's core, leaving no HSE signature in the mantle. However, by considering a collisional size distribution, we show that to deliver sufficient mass in small impactors to account for Earth's HSEs, there will be an implausibly large mass delivered by larger bodies, the metallic fraction of which lost to Earth's core. There is therefore a contradiction between observed concentrations of HSEs, the geodynamics of metal entrainment, and estimates of total mass accretion during the late veneer. To resolve this paradox, and avoid such a mass accretion catastrophe, our results suggest that large impactors must contribute to observed HSE signatures. For these HSEs to be entrained in the mantle, either some mechanism(s) must efficiently disrupt impactor core material into $\leq\,$0.01\,mm fragments, or alternatively Earth accreted a significant mass fraction of oxidised (carbonaceous chondrite-like) material during the late veneer. Estimates of total mass accretion accordingly remain unconstrained, given uncertainty in both the efficiency of impactor core fragmentation, and the chemical composition of the late veneer.
\end{abstract}

\section*{Plain Language Summary}
Highly siderophile elements (HSEs) have a very strong tendency to partition into planetary cores, rather than mantles. If Earth's mantle and core were chemically equilibrated, these elements should be almost non-existent in the mantle. Yet, these elements are much more abundant in the mantle than expected, and are present in roughly the same relative abundance as in chondritic meteorites. A widely-admitted hypothesis is that these elements were delivered as a `late veneer' of chondritic material, carrying about $0.5\,$\% of Earth's mass after core formation was complete. This estimate assumes that all HSEs delivered during the late veneer remained suspended in Earth's mantle. In this work, we show that it is very challenging for these elements, delivered by planetesimals larger than approximately 1\,km, to avoid sinking to Earth's core, due to the large density of these metals relative to Earth's silicate mantle. Our calculations further show, by considering a realistic planetesimal size distribution, that there is insufficient mass in small planetesimals to account for Earth's HSEs. These results therefore highlight a contradiction between estimates of mass accretion during the late veneer, and our understanding of metal delivery to Earth's mantle.

%
%

\section{Introduction}
\label{sec:introduction}

Highly siderophile elements (HSEs; namely Pt-group elements, Re and Au) have a very strong affinity for Fe-metal at low pressure, and should therefore be stripped from the molten silicate mantle during core formation. Partitioning data predict that HSEs should be both virtually non-existent in Earth's mantle, and, in those that remain, chemically fractionated according to their different affinities for metal \cite{Righter2008, Mann2012}. In contrast to these two predictions, the Earth's mantle contains a significant excess of HSEs compared to that expected from metal-silicate equilibration \cite{Day2007, Walker2009}, and these HSEs are found to be in nearly chondritic relative abundance \cite{Day2016}. The simplest explanation for these observations is the delivery of a `late veneer' of chondritic material, with total mass $\sim\,$0.3$-$0.7\,\% of the Earth \cite{Kimura1974, Chou1978, Walker2009}, overprinting the mantle's post-core formation HSE composition. 

Further evidence in support of a significant, and widespread period of late accretion is that excess HSEs are also found, in broadly chondritic relative abundance, in the mantles of \add{much smaller planetary bodies including} the Moon \cite{Day2007}, Mars \cite{Brandon2000, Brandon2012}, and Vesta \cite{Dale2012}. This strongly indicates that high-pressure equilibration does not alone control observed concentrations of HSEs in these bodies\add{, since it appears unlikely that pressure-temperature conditions were the same at the bottom of these respective magma oceans}. 
Thus, despite measurements indicating that the affinity of HSEs for metal decreases in high pressure-temperature conditions \cite{Righter2008, Mann2012, Suer2021}, a chondritic late veneer still appears necessary in order to account for the abundance of HSEs observed in the terrestrial planets' mantles.

The Earth-Moon system provides a particularly sensitive test for models of late accretion, given both experienced early global differentiation \cite{Kleine2005, Boyet2005}, and shared a common impact bombardment. A consistent explanation for both the Earth, and Moon's HSE inventories remains enigmatic however, given that lunar HSEs are found in chondritic relative proportions, but with a concentration 20$-$40 times lower than estimates for the terrestrial mantle \cite{Day2007, DayWalker2015}. Most straightforwardly, this concentration would imply that total mass accretion to the Earth was three orders of magnitude greater than to the Moon, a discrepancy that cannot be attributed solely to the larger geometric cross section of the Earth \cite <e.g.,>{Walker2009}.

Several quite different explanations have been subsequently proposed for this dearth of lunar HSEs. \citeA{Bottke2010} first suggested the majority of Earth's HSEs were delivered by several large ($D\,>\,2000\,$km), leftover planetesimals from a population dominated by large objects. These large bodies would be more likely to be accreted by the Earth given its larger (gravitationally enhanced) accretional cross section, thus explaining the large discrepancy in Earth-Moon HSE abundances. This was supported on dynamical grounds by \citeA{Raymond2013}, who showed that by assuming a low initial angular momentum deficit, a late veneer dominated by large bodies could reproduce both Earth's late accreted mass, and the current-day orbital excitation of the terrestrial planets. Not only this, but the large Earth-Moon HSE abundance ratio could also be accounted for, given that the erosional nature of large impacts on the Moon naturally prevents the accretion of large ($>\,500\,$km) bodies \cite{Raymond2013}. Later work suggested that Earth's HSEs might instead have been delivered by a single lunar-sized impactor, rather than multiple Ceres-sized objects \cite{Brasser2016}.

Alternatively, \citeA{Schlichting2012} proposed a late veneer of very small ($\sim\,$10\,m) planetesimals, which are collisionally damped to very low eccentricities, thereby increasing the relative gravitational focussing ratio in favour of the Earth. These planetesimals, assumed to form small \cite <e.g.,>{Weidenschilling2011}, would naturally explain the damping of the terrestrial planets' eccentricities and inclinations following giant impacts. However, the extent to which such a population of small planetesimals could remain on near-circular orbits and avoid re-excitation by the terrestrial planets is unclear, and there is no alternative evidence supporting such a collisionally damped disk in the inner Solar System.

These studies all assume that HSEs are only removed from planetary mantles during core formation, and that their present abundances reliably trace subsequent mass accretion. Estimates of total mass accretion during the late veneer therefore strongly depend on this assumption. HSEs can, however, be removed from the mantle post-core formation; for example, the significant re-melting of the upper mantle during large impacts \cite <e.g.,>{Nakajima2021} has the capacity to strip extant HSEs delivered by previous (smaller) impactors.

The assumption that mantle HSE abundances reliably trace total mass accretion appears, however, questionable from a mineral physics perspective.
\citeA{Rubie2016} highlighted that HSEs could be removed from the mantle via the exsolution, and segregation of iron sulfide, which may occur during the crystallization of planetary magma oceans. \citeA{Rubie2016} therefore proposed the alternative interpretation that HSEs only record mass accretion post-magma ocean crystallisation, which may occur much later than the initial stage of core-mantle differentiation. This \remove{may}\add{would} have significant consequences for the Earth-Moon system, given the lunar magma ocean may have crystallised up to 200\,Myr after the Earth's, due to \remove{strong tidal heating, and} the \add{rapid} formation of an insulating anorthositic crust \cite{Elkins-Tanton2008, ElkinsTanton2011}. \add{While recent works suggest it is unlikely that the lunar (or terrestrial) magma ocean reached sulfur saturation \protect\cite{Steenstra2020,Blanchard2025}, both delayed lunar core formation \cite{DayPaquet2021}, and tidally driven remelting of the lunar mantle \cite{Nimmo2024} would similarly strip extant HSEs from the lunar mantle.}
It is therefore possible, given the steep decline in impact rate at early times, that the Moon accreted significantly more mass than appears to be recorded by its HSE record \cite{Morbidelli2018}.

\begin{figure}
    \centering
    \includegraphics[width=\textwidth]{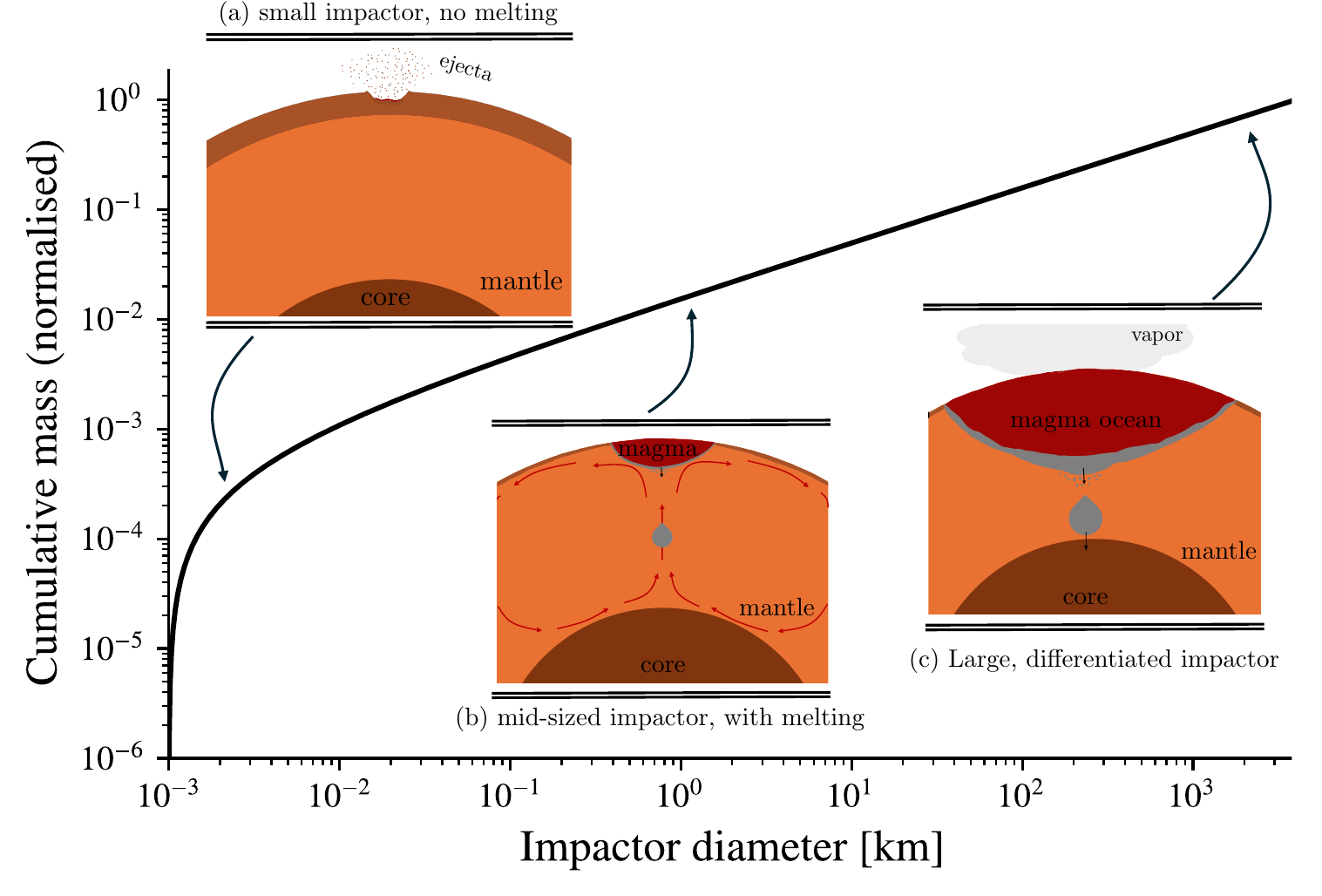}
    \caption{
    Collisions between leftover planetesimals drives the redistribution of mass between bodies of different sizes, generating a so-called collisional size-frequency distribution (SFD) dominated, in number, by the smallest bodies. The Earth will therefore, unavoidably, accrete planetesimals of a wide range of sizes during the late veneer. For the canonical collisional SFD \cite{Dohnanyi1969}, total mass is concentrated in the largest planetesimals (plotted above). The geodynamical processes responsible for the delivery of HSEs to the mantle, are controlled, primarily, by the size of the impactor, and therefore have the capacity to dramatically bias estimates of total mass accretion during the late veneer. We identify three possible regimes of HSE delivery, with illustrative schematic diagrams inset above.
    (a) Small impactors at low velocity will generate little melt, and are expected to fragment into millimetric pieces (see \S\ref{sec:post_impact_HSE_distribution}). The ability of these impactors to affect mantle geochemistry will depend on whether the tectonic regime of the planet enables them to be recycled into the mantle. (b) Small impactors at high velocity will generate significant melt, from both the target and impactor \cite{Melosh1989}. We expect metal diapirs to quickly enter the solid mantle (\S\ref{sec:metal_entrainment_magma_pond}), bringing with them the impactor's HSEs. The ability of these impactors to affect mantle geochemistry will depend on whether the frictional force resisting diapir descent is larger than the negative buoyancy. (c) Large, differentiated impactors will generate large volumes of melt. Unless impactor core material can be fragmented into very small droplets, large diapirs will enter the solid mantle, and quickly sink to Earth's core (\S\ref{sec:metal_entrainment_mantle}).
    }
    \label{fig:HSE_accretion_schematic}
\end{figure}

The assumption that all delivered HSEs remain in Earth's mantle is also questionable from a geodynamical perspective. This is particularly problematic if the majority of mass accreted during the late veneer was delivered by large, differentiated planetesimals \cite <e.g.,>{Bottke2010, Brasser2016, Morbidelli2018}. With the metallic core of such a differentiated planetesimal twice as dense as the silicate mantle, a significant fraction of this metal will likely sink through the mantle under the action of gravity, and eventually merge with Earth's core \cite <e.g.,>{Rubie2003, Deguen2014, Clesi2020}. Thus, a significant fraction of this metal will be lost to the core without contributing to the mantle HSE signature of an impact \cite <e.g.,>{Marchi2018}. 
Estimates of total mass accretion during the late veneer therefore depend crucially on the size distribution of leftover planetesimals, and the corresponding geodynamical processes controlling the efficiency of HSE delivery to the mantle (see figure~\ref{fig:HSE_accretion_schematic}).

The late accretion of large, differentiated planetesimals might therefore \add{appear to} be an inefficient mechanism for delivering metals to Earth's mantle. This is, however, in tension with strong chemical and isotopic constraints, which support the addition of impactor core material to Earth's mantle \cite{Kleine2009, DahlStevenson2010}. Several recent studies have investigated the fate of impactor core material using both SPH and hydrocode simulations, all finding significant mass loss to Earth's core \cite <e.g.,>{Genda2017, Marchi2018, CitronStewart2022, Itcovitz2023}. The proposed mechanisms of HSE delivery to the mantle however differ, and include the elongation and disruption of lunar-sized embryos \cite{Genda2017}, rapid three-phase flow at the base of an impact-generated magma ocean \cite{KorenagaMarchi2023}, and the vaporisation of impactor core material \cite{Albarede2013, Kraus2015, Itcovitz2023}. Many important processes, dictating the fate of impactor core material, remain however on length scales far below simulation resolution. This is particularly challenging for many sources of turbulence, given that responsible instabilities are unable to propagate onto length scales that can be resolved \cite <e.g.,>{DahlStevenson2010, Deguen2014}. Accordingly, it is very challenging to accurately constrain the extent of mass loss to Earth's core.

In this study we combine both geophysical and astrophysical arguments to constrain the origin of Earth's mantle HSEs, and demonstrate that there is a contradiction between the observed concentrations of HSEs in the mantle, the geodynamics of HSE delivery, and current estimates of total mass accretion during the late veneer. In \S\ref{sec:entrainment_methods} we motivate the size-dependence of efficient HSE delivery, and determine a critical impactor size below which HSEs are efficiently entrained in the mantle. We present an analytical discussion of the buoyancy of metallic iron in both an impact-generated magma pond and a solid mantle, demonstrating that HSEs from impactors larger than approximately 1\,km will be lost to Earth's core. We investigate in \S\ref{sec:small_impactors_methods} the ability of small impactors to deliver observed HSEs, calculating the total mass accretion from larger bodies in an asteroid-like size distribution. In \S\ref{sec:discussion} we discuss the implications of this study for HSE delivery to the Earth and Moon, and for estimates of total mass accretion during the late veneer.

\section{The entrainment of metal and its HSEs in the mantle}
\label{sec:entrainment_methods}

In this section, we demonstrate the importance of impactor size in determining the fate of its HSEs. In \S\ref{sec:post_impact_HSE_distribution} we discuss the fate of an impactor's metal, and its HSEs, in the immediate aftermath of an impact. We then discuss the post-impact dynamics, as the impactor's metal experiences a negative buoyancy force in an impact-generated magma pond (\S\ref{sec:metal_entrainment_magma_pond}), and a solid mantle (\S\ref{sec:metal_entrainment_mantle}). From these discussions, we determine the critical impactor diameter above which an impactor's HSEs are lost to the core, and do not contribute to observed mantle HSE signatures.

\subsection{The dispersal of metal and HSEs in the aftermath of an impact}
\label{sec:post_impact_HSE_distribution}

\add{Immediately after the Moon-forming impact, Earth's mantle will have been predominantly (if not fully) molten \protect\cite <e.g.,>{Canup2008,NakajimaStevenson2015,Lock2018}. Unlikely to have formed an insulating, conductive lid \cite{Elkins-Tanton2008}, mantle solidification will have been regulated by the transfer of heat through Earth's primitive atmosphere, occurring within $\sim\,5$\,Myr of Moon-formation \cite{Elkins-Tanton2008,Hamano2013}. While the duration of late accretion remains uncertain, dynamical models suggest that the depletion timescale of leftover planetesimals was an order of magnitude larger \cite{Morbidelli2018,Brasser2020}. We therefore assume throughout that Earth's mantle had solidified post-Moon formation.}

The ability of an impactor to contribute to mantle HSEs \remove{depends}\add{will depend} on the spatial distribution of the impactor's metals immediately after an impact. This, in turn, is largely controlled by the impactor's initial structure (i.e., whether or not it is differentiated), its fate during collision with the Earth, and its ability to reach and melt the Earth's mantle. \add{This last point is of particular importance, given that the geodynamics of metal descent through Earth's mantle depends sensitively on mantle viscosity \protect\cite{DahlStevenson2010, Samuel2012}. Since the viscosity of impact-generated melt will be more than 10 orders of magnitude smaller than the solid (crystallized) mantle below \cite{Solomatov2015}, the impact-driven melting of Earth's mantle is expected to play a central role in controlling the efficiency of HSE delivery.}

We therefore anticipate a large diversity of post-impact scenarios (see figure~\ref{fig:HSE_accretion_schematic}), with substantially different consequences for the partitioning of HSEs between the mantle and core. 
\remove{We expect this diversity is largely driven by the impactor's size, as this predominantly controls the specific impact energy.} 
\add{\protect We expect this diversity to be largely driven by impactor size, given this is expected to vary by several orders of magnitude for realistic size distributions (see figure~\ref{fig:HSE_accretion_schematic}), and will therefore predominantly control the specific impact energy.}
This is illustrated most clearly through comparison of small impactors (figure~\ref{fig:HSE_accretion_schematic}a), which will generate little melt and remain embedded in the Earth's crust, and large differentiated impactors (figure~\ref{fig:HSE_accretion_schematic}c), which will generate significant silicate melt extending into the convective mantle.
We discuss these end-member scenarios first, before addressing the transition (at intermediate sizes) between the efficient and inefficient delivery of HSEs to the mantle.

Evidence from terrestrial impact craters, which record little melting of the target, indicate that small impactors are fragmented into millimetric pieces during collisions with the Earth \cite{Blau1973, Melosh2005, Folco2022}, which may be subsequently oxidised by Earth's hydrosphere. The ability of small impactors to contribute to mantle geochemistry is therefore entirely reliant on the recycling of Earth's early crust. While there is debate regarding the Hadean Earth's tectonic regime, it is very likely that this early crust was recycled relatively quickly during the late veneer \cite <e.g.,>{RosasKorenaga2018}. We will later demonstrate, in \S\ref{sec:metal_entrainment_mantle}, that these millimetric fragments are small enough to remain in the solid target, and thus small impactors are able to efficiently contribute to Earth's HSEs, albeit over geologic timescales.

In contrast, the impacts of large differentiated bodies are violent events, which will readily break through the crust, melt the outer part of the mantle, and generate large volumes of silicate melt \cite{TonksMelosh1993, Nakajima2021}. The impact-generated shock wave will similarly melt the impactor, such that its metallic liquid core is released into the magma pond; these metal droplets are more than twice as dense as the surrounding molten silicates, and are thus liable to quickly sink, and collect at the bottom of the newly-formed magma pond \cite{Deguen2014}. The geodynamics of this process, and the subsequent fate of metals as they enter the convective mantle are therefore crucial to an impactor's ability to contribute to mantle geochemistry. This is discussed next, in \S\ref{sec:metal_entrainment_magma_pond} and \S\ref{sec:metal_entrainment_mantle}, in which we demonstrate that a significant fraction of impactor core material will be lost to Earth's core.

We therefore anticipate that there must exist a critical impactor diameter, $D_{\rm crit}$, which delineates the transition between the fragmentation of smaller impactors (which will distribute their HSEs throughout the Earth's crust), and the generation of significant melt during much larger impacts (with their metals subsequently liable to gravitational instability, and loss to Earth's core). In the following paragraphs, we estimate the value of this critical, intermediate, impactor diameter, $D_{\rm crit}$.

Through comparison of the above end-member scenarios, it is apparent that the extent of melting during an impact is fundamentally important in determining the post-impact fate of metals. Previous studies have demonstrated that the amount of impact-generated melt depends on both the impact velocity and angle \cite <e.g.,>{Pierazzo1997}, which determine the peak shock pressure experienced during compression. However, this is very complex for oblique impacts \cite{PierazzoMelosh2000}. Using the results from \citeA{PotterCollins2013}, who studied this in detail, we find the critical impact velocity for \remove{significant melting} \add{the significant melting of impactor material} is in the range 12$-$15\,km\,s$^{-1}$ (see~\ref{sec:appendix_melting}). Such high velocity impacts will concurrently generate significant melting of the target, which requires $v_{\rm imp}^2/E_m\gtrsim30$, where $E_m$ is the internal energy of melting \cite{Pierazzo1997}.

The impact velocity of small bodies at Earth's surface, $v_{\rm imp}$, depends on both the entry speed of an impactor (which will always exceed 11.2\,km\,s$^{-1}$), and its interaction with the atmosphere. The significance of this interaction, dominated by atmospheric deceleration and fragmentation, is largely governed by the impactor's size (with large bodies able to reach the surface almost unperturbed), and the surface density of Earth's early atmosphere \cite <e.g.,>{Chyba1993, Svetsov1995, Melosh2005}. The evolution of Earth's atmosphere remains subject to debate, particularly during the Hadean, but is thought to have been initially very dense ($\sim\,$100\,bar), following the degassing of Earth's magma ocean \cite{Elkins-Tanton2008}. A relatively quick transition in atmospheric pressure is then expected, which by the end of the Hadean is inferred to be less than 1\,bar \cite{Rimmer2019, CatlingZahnle2020}. Using the simple atmospheric entry model from \citeA{Chyba1993}, we find the critical diameter for melting thus varies in the range $\sim\,$10$-$1000\,m (see figure~\ref{fig:critical_impactor_size_melting})\add{, corresponding to the plausible variation in atmospheric surface pressure during the Hadean (see \protect\ref{sec:appendix_melting} for more detailed discussion of this model)}. Smaller impactors will be unable to reach the surface intact, or at a sufficiently high velocity as required \remove{for melting} \add{to melt both impactor and target material}. This suggests the critical diameter, $D_{\rm crit}$, is at most on the order of 1\,km.

Another important consideration, that is likely to inform the value of the critical diameter $D_{\rm crit}$, is the ability of an impactor to reach the convective mantle. When the impactor's metal is directly implanted into the convective mantle, it will immediately sink due to its density difference with the surrounding silicates, and can be potentially lost to the core (figure~\ref{fig:HSE_accretion_schematic}). In contrast, metals injected into the crust will remain trapped there until admixed into the mantle. In the latter scenario, metals may be oxidised before reaching the mantle, and hence may contribute to observed HSE signatures. The minimum impactor size required to penetrate Earth's crust can be estimated using scaling laws for impact crater depths. Using the crater depth scaling from \citeA{Allibert2023}, we find that for a 1\,km diameter impactor (assuming a typical impact velocity of 20\,km\,s$^{-1}$), crater depths exceed 10\,km, which is sufficient to penetrate modern oceanic crust. Whilst the thickness of Earth's early crust is uncertain and subject to significant debate \cite{Korenaga2018}\remove{, it is}\add{, it was likely much thinner than of the modern Earth.}
\add{It therefore appears overwhelmingly} likely \remove{1\,km impactors}\add{that impactors larger than 1\,km} would penetrate Earth's early crust.

Given the consensus between these two independent lines of investigation, we therefore suggest that $D_{\rm crit}\sim1\,$km marks an important transition in the post-impact fate of HSEs; smaller bodies will fragment into small pieces, distributing their HSEs throughout Earth's crust, whereas larger bodies will simultaneously penetrate the crust, and generate significant melt. The true fate of HSEs will of course be the result of many physical processes spanning a wide range of length scales; we therefore stress there remains uncertainty in this critical impactor size, which we revisit in \S\ref{sec:methods_limitations}. Next, we focus on the fate of larger impactors ($D>1\,$km), first investigating the fate of metals in a magma pond (\S\ref{sec:metal_entrainment_magma_pond}), and then the solid mantle (\S\ref{sec:metal_entrainment_mantle}).

\subsection{The entrainment of metal in a magma pond}
\label{sec:metal_entrainment_magma_pond}

\add{During collision with Earth, high-resolution simulations \protect\cite <e.g.,>{KendallMelosh2016} observe the stretching of an impactor's metallic core, which is spread into a thin layer over the crater floor. This metal is stretched further during the formation of a strong, vertical jet during crater collapse, fragmenting $\sim\,100\,$km cores into km-scale `blobs' \cite{KendallMelosh2016} -- the resolution limit of these simulations. Analogue experiments reveal that turbulent mixing drives further metal fragmentation down to the capillary scale, forming small droplets stabilized via surface tension \cite{Deguen2014, WacheulLeBars2018, Landeau2021, Maller2024}. Despite the fast equilibration expected between these small metal droplets and the silicate melt \cite{Ulvrova2011,LhermDeguen2018,Clesi2020}, we expect there will be no significant flux of elements between these phases given the large metal-silicate partition coefficients of the HSEs \cite{Righter2008,Mann2012}. Hence, we restrict our attention to the entrainment of impactor metal, and its HSEs, in the newly-formed magma pond.}

\remove{We assume that, after a large impact, the impactor's metal fragments into drops of diameter $d$ \protect\cite{Deguen2014, WacheulLeBars2018, Maller2024}.} \remove{These metal drops, of density $\rho_m$, then} 
\add{We assume that the impactor's metal core fragments into drops of diameter $d$ and density $\rho_m$, which} settle in a fully liquid silicate magma pond of viscosity $\mu_s$, density $\rho_s$, and depth $H$, \remove{which convects} \add{convecting} turbulently at velocity $U$.
Particles can be suspended in a convective layer when the frictional force is large enough relative to the buoyancy force of each particle \cite{Solomatov1993, Sturtz2021, Monteux2023}. The Shields number 
\begin{linenomath*}
    \begin{equation}
        \label{eq:shields_number}
        \theta_S = \frac{\tau}{\Delta \rho\, g\, d},
    \end{equation}
\end{linenomath*}
compares these two forces, where $\tau$ is the shear stress induced by the convection, $\Delta\rho=\rho_m-\rho_s$ the density difference, and $g$ the gravitational acceleration. 

Previous studies have shown that there exists a critical value, above which particles can be re-entrained by convective motions, which is given by $\theta_c=0.15\pm0.05$ \cite{Solomatov1993, Sturtz2021, Monteux2023}. Drops with $\theta_S<\theta_c$ will always sediment out of the convecting layer. While this condition has only been tested for solid particles, we expect that the \remove{surface tension of liquid drops}\add{additional stabilising force provided by surface tension} will make \remove{their}\add{the} entrainment \add{of liquid drops} even more challenging.

In a fully liquid magma pond with a depth, $H$, of $\sim$\,1000\,km, convective speeds are typically on the order of $10\,{\rm m\,s}^{-1}$ \cite{Solomatov2015}. With a low viscosity on the order of $0.05$\,Pa\,s \cite{StixrudeKarki2010}, the Reynolds number, which measures the ratio of inertia to viscous forces, is larger than $10^{11}$, meaning that the flow is fully turbulent \add{\protect\cite{SalvadorSamuel2023}}. In such a turbulent magma pond, the largest shear stresses are likely the Reynolds stresses \cite{Solomatov1993,Solomatov2015},
\begin{linenomath*}
    \begin{equation}
        \label{eq:tau_turbulent}
        \tau = \rho_s\,u^2, 
    \end{equation}
\end{linenomath*}
where $u$ is the friction velocity \cite{Shraiman1990},
\begin{linenomath*}
    \begin{equation}
        \label{eq:friction}
        u=\dfrac{U}{x},
    \end{equation}
\end{linenomath*}
and $x$ satisfies 
\begin{linenomath*}
    \begin{equation}
        \label{eq:friction_x}
        x = 2.5\,\log{\left(\frac{\rho_s \, U\,H}{\mu_s} \dfrac{1}{x}\right)}+6.
    \end{equation}
\end{linenomath*}

We assume that the convective velocity $U$ follows the scaling for hard turbulence \cite{Shraiman1990,Solomatov2015}
\begin{linenomath*}
    \begin{equation}
        \label{eq:ConvectiveSpeed_Turbulent}
        U = 0.086\, x\, \left(\dfrac{\alpha\,g\,H\,F}{\rho_s\,C_p}\right)^{1/3},
    \end{equation}
\end{linenomath*}
where $\alpha$ is the thermal expansion coefficient, $F$ the heat flux at the top of the magma pond, and $C_p$ the specific heat capacity under constant pressure. Using the typical values for $U$, $\mu_s$, and $H$ in a deep magma ocean (Tables~\ref{tab:properties}~and~\ref{tab:properties_range}), one finds that $x$ is in the range $60-75$, and the friction speed is typically in the range $0.1-0.5$\,m\,s$^{-1}$.

\begin{table}[t]
    \centering
    \begin{tabular}{c|ccccccc}
        \toprule
         Parameter & $g_\oplus$ & \remove{$g$, Moon}\add{$g_L$} & $\alpha$ $^a$& $\rho_s$ $ ^b$ & $\rho_m$ $^c$& $C_p$  $^d$ & $k_s$  $^e$  \\
         Unit & m\,s$^{-2}$ & m\,s$^{-2}$ & K$^{-1}$  & kg\,m$^{-3}$ & kg\,m$^{-3}$&  J\,kg$^{-1}$ & m$^2$\,s$^{-1}$  \\
         \midrule
        Value & $9.8$ & $1.62$ & $3\times 10^{-5}$ & $4500$ & $9000$ & $10^3$ &$10^6$  \\
        \bottomrule
    \end{tabular}
    \caption{Assumed mantle and metal properties used in sections \ref{sec:metal_entrainment_magma_pond} and \ref{sec:metal_entrainment_mantle}. Values are compatible with published studies: $^a$ \cite{Chopelas1992}, $^b$ \cite{Miller1991}, $^c$ \cite{Morard2013}, $^d$ \cite{Stebbins1984}, $^e$ \cite{deKoker2010}.}
    \label{tab:properties}
\end{table}

\begin{table}[t]
    \centering

    \begin{tabular}{c|ccccc}
        \toprule
         Parameter & $\Delta T$ & H & $U$ & F & $\mu_s$ \\
         Unit & K & m & m\,s$^{-1}$ & W\,m$^{-2}$ & Pa\, s \\
         \midrule
        Value & $500-2500$ &$10^6-3\times10^6$ & $4-40$ & $3\times 10^{5}-10{^6}$& $0.02-0.08$ \\
        \bottomrule
    \end{tabular}
    
    \caption{Ranges of plausible parameter values relevant for a fully liquid magma ocean. Ranges for $\Delta T$, $H$ and $U$ are from \cite{Solomatov2000}, the viscosity range from \cite{StixrudeKarki2010} and the range of possible heat flux $F$ from several studies \cite{Solomatov2015,Lebrun2013}.} 
    \label{tab:properties_range}
\end{table}

\begin{figure}[t]
    \centering
    \includegraphics[width=\textwidth]{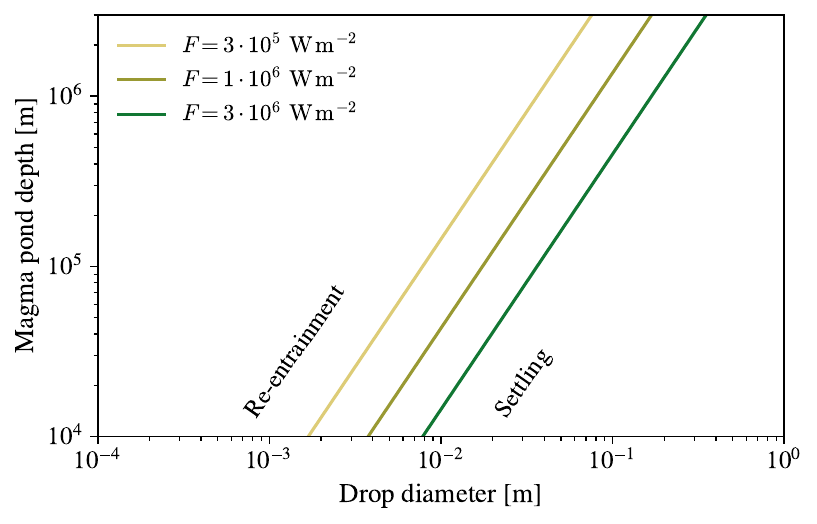}
    \caption{The maximum diameter of metal drops that can be entrained by turbulent convection in a fully liquid magma pond, $d_{\text{entr.}}$ (corresponding to the condition $\theta_S=\theta_c$; equation \ref{eq:crit_radius_turb}), as a function of magma pond depth for three plausible values of the heat flux $F$. The friction speed $u$ in equation~(\ref{eq:crit_radius_turb}) is computed from equations \ref{eq:friction}-\ref{eq:ConvectiveSpeed_Turbulent}. Note that $\theta_S=\theta_c$ is a necessary, not sufficient condition for entrainment, and so metal drops smaller than this maximum diameter will not all remain suspended in equilibrium (see equation~\ref{eq:entrained_fraction}, figure~\ref{fig:suspended_metal_MO_Earth}).}
    \label{fig:CriticalDiapirRadius_MO}
\end{figure}

Inserting (\ref{eq:tau_turbulent}) into the definition of the Shields number (\ref{eq:shields_number}), one finds that the condition $\theta_S>\theta_c$ is met when the drop diameter is smaller than 
\begin{linenomath*}
    \begin{equation}
        \label{eq:crit_radius_turb}
        d_{\text{entr.}} = \dfrac{\rho_s \, u^2}{\theta_c \, \Delta \rho \, g}.
    \end{equation}
\end{linenomath*}
With a friction speed of about 0.1$-$0.5\,m\,s$^{-1}$, the critical diameter $d_{\text{entr.}}$ for entrainment is on the order of $1-10$\,cm; drops smaller than $1$\,cm may therefore be entrained by the convection. This critical drop size is for a deep magma pond on the order of $1000\,$km. In shallower magma ponds, the convective velocity (equation~\ref{eq:ConvectiveSpeed_Turbulent}), the friction speed (equation~\ref{eq:friction_x}), and hence the Reynolds stresses (equation~\ref{eq:tau_turbulent}) are lower, making the entrainment of metal drops significantly harder. This is shown in figure~\ref{fig:CriticalDiapirRadius_MO}, where we observe that the maximum drop diameter for entrainment decreases as the magma pond depth decreases.

The diameter $d$ of metal drops in magma oceans remains poorly known. Yet, previous studies have obtained orders of magnitude estimates for the drop size when metal fragments after a large planetary impact \cite{Deguen2014}. Using existing theories for fragmentation in a turbulent flow \cite{Hinze1955}, they obtain $d\sim 1$~mm \cite{Deguen2014}. More recent experiments on the impact of a liquid onto an immiscible liquid bath also suggest that the impactor core fragments into drops of 0.3\,mm in Earth's magma ocean \cite{Maller2024}. Because these drops are smaller than $d_{\text{entr.}}\sim 1$~cm, they will not necessarily sediment directly out of the convecting layer.

The condition $\theta_S>\theta_c$ is, however, a necessary, but not sufficient condition for the entrainment of metal drops. When $\theta_S>\theta_c$, drops will constantly sediment out of the convective region, whilst others are re-entrained by the convective motions \cite{Solomatov1993}. When the flux of settling drops balances the flux of re-entrained ones, the mass of suspended drops reaches an equilibrium. The mass of drops kept in suspension is then dictated by an energy balance \cite{Solomatov1993b}. Indeed, only a fraction $\epsilon$ of the energy generated by buoyancy forces is converted into the gravitational energy of the dense, suspended metal drops. The suspended (volume) fraction of metal drops, at equilibrium, then reads \cite{Solomatov1993b},
\begin{linenomath*}
    \begin{equation}
        \label{eq:entrained_fraction}
        \Phi = \epsilon \, \dfrac{\alpha \, F}{C_p \, \Delta \rho \, U_d } 
    \end{equation}
\end{linenomath*}
where $\epsilon$ is an empirical, dimensionless efficiency factor that varies in the range $0.2\%-0.9\%$ \cite{Solomatov1993,LavorelLeBars2009}, $U_d$ the settling speed of an individual drop. 

We assume that the settling speed is given by,
\begin{linenomath*}
    \begin{equation}
        \label{eq:settling_speed}
        U_d = \sqrt{\frac{4\,\Delta \rho\,g\,d}{3\,\rho_s\,C_d}},
    \end{equation}
\end{linenomath*}
where $C_d$ is the drag coefficient. The value of $C_d$ depends on the drop Reynolds number $Re_d = \rho_s\, U_d \, D/\mu_s$, which measures the relative importance of inertial and viscous forces around the drop. The value of $Re_d$ is typically lower than $1$ for metal drops with a diameter $d\lesssim 10^{-4}$~m but it reaches $10^3$ for larger drops with $d=1$\,cm. In this range of $Re_d$, the settling velocity $U_d$ varies from the Stokes regime, when $Re_d < 1$, to a regime of nearly constant drag coefficient. To describe this transition we use the empirical model proposed by \cite{Samuel2012} in which
\begin{linenomath*}
    \begin{equation}
        \label{eq:drag}
        C_d = \dfrac{24}{Re_d}+c_N,
    \end{equation}
\end{linenomath*}
where $c_N=0.3$, $Re_d = \rho_s\, U_{d,S}\, d/\mu_s$ and 
\begin{linenomath*}
    \begin{equation}
        \label{eq:Stokes}
        U_{d,S}=\dfrac{g\, \Delta \rho d^2}{18\, \mu_s}
    \end{equation}
\end{linenomath*}
is the Stokes velocity.

With this choice, when $Re_d \ll 1$, $C_d \simeq 24/Re_{d,S}$ and $U_d$ from (\ref{eq:settling_speed}) tends to the Stokes velocity (\ref{eq:Stokes}), which increases as the square of the drop diameter. Instead, in the limit of large $Re_d$,  $C_d$ tends to the value $c_N=0.3$, meaning that $U_d$ increases as the square root of the drop diameter.

\begin{figure}
    \centering
    \includegraphics[width=1\textwidth]{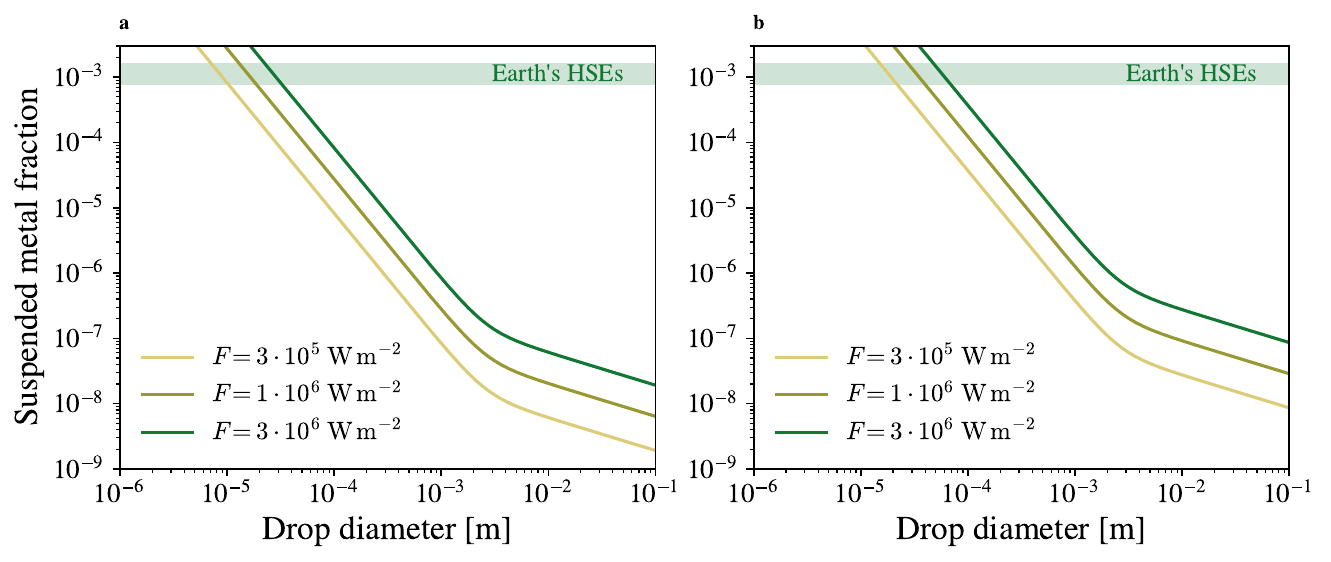}
    \caption{Volume fraction of suspended metal in a turbulent magma ocean on an Earth-sized planet as a function of the metal drop diameter, for three plausible values of the heat flux $F$. The green shaded band locates the suspended volume fraction needed so that the HSE concentration in the magma pond matches that of the present-day Earth's mantle. We use equations~\ref{eq:entrained_fraction}-\ref{eq:Stokes} assuming $g=9.8\,{\rm m\,s}^{-2}$, $\rho_m=9000\,{\rm kg\,m}^{-3}$, $\rho_s=4500\,{\rm kg\,m}^{-3}$, $\mu_s=0.05$\,Pa\,s and $H = 2000\,$km. (a) Lower-end estimates assume \remove{$\epsilon=0.3\%$}\add{$\epsilon=0.2\%$}. (b) Upper-end estimates with $\epsilon=0.9\%$.}
    \label{fig:suspended_metal_MO_Earth}
\end{figure}

Inserting equations~(\ref{eq:settling_speed}), (\ref{eq:drag}), and (\ref{eq:Stokes}) into equation~(\ref{eq:entrained_fraction}), one obtains the volume fraction $\Phi$ of entrained metal as a function of the drop diameter $d$. In all regimes, the larger the drops, the larger the settling speed $U_d$, and hence, the lower the fraction of entrained metal $\Phi$. This suspended fraction is shown in figure \ref{fig:suspended_metal_MO_Earth} as a function of the drop diameter $d$ for an Earth-like planet with a deep magma ocean, and for different values of the heat flux $F$. 
\remove{In a fully liquid magma pond, the heat flux at the surface of the magma ocean, as predicted by 1D thermal evolution models, is in the range $3\times 10^{5}-10^{6}$\,W\,m$^{-2}$ \protect\cite{Solomatov2000, Lebrun2013, Salvador2017}.}
\add{In a fully liquid magma pond, with viscosity $\lesssim 0.1\,{\rm Pa\,s}$, 1D convective thermal evolution models predict a surface heat flux in the range $3\times 10^{5}-10^{6}$\,W\,m$^{-2}$ \protect\cite{Solomatov2000, Lebrun2013, Salvador2017}.}
To be conservative, we explore the larger range $3\times 10^{5}-3\times10^{6}$\,W\,m$^{-2}$. 

Assuming millimetric metal drops, the volume fraction of suspended metal is more than three orders of magnitude smaller than required to account for observed HSEs in Earth's mantle (figure \ref{fig:suspended_metal_MO_Earth}). We predict that, to keep enough metal in suspension and match HSE concentrations in Earth's mantle, the metal drops need to be smaller than $3\times10^{-5}$\,m with $F=10^6$\,W\,m$^{-2}$ (figure \ref{fig:suspended_metal_MO_Earth}). Even with an extreme heat flux of $3\times 10^6$\,W\,m$^{-2}$, drops need to be smaller than $5\times10^{-5}$\,m. \remove{This is at least one order of magnitude smaller than the drop size expected after an impact (Deguen et al., 2014; Maller, 2024).} \add{This is more than one order of magnitude smaller than the mean drop size expected after an impact \protect\cite <$\sim1\,$mm;>{Deguen2014,WacheulLeBars2018,Maller2024}. While there will be a distribution of drops below this size, much of the mass expected to be in the largest drops. There will therefore be insufficient mass in small drops to account for Earth's HSEs (see \ref{sec:appendix_drop_distribution}).
}

We expect this critical drop size to be independent of the depth of the magma pond ($H$), given that the suspended metal fraction is proportional to the heat flux at the surface of the magma pond ($F \propto Ra^{1/3} / H$), which is itself independent of $H$ given that the Rayleigh number scales as $Ra \propto H^3$. We further note that the suspended metal fraction changes insignificantly if we assume a lower density $\rho_m \approx 7800\,{\rm kg\,m}^{-3}$, as may be appropriate for smaller magma ponds. Unless additional fragmentation mechanisms generate drops smaller than $0.05\,$mm, we therefore expect metal (from molten impactors) to rapidly settle and coalesce into a layer at the bottom of the newly-formed magma pond. In \S\ref{sec:metal_entrainment_mantle} we consider the stability of these metal pools (which overlay a layer of solid, or partially solid mantle; see figure~\ref{fig:HSE_accretion_schematic}b), and the subsequent fate of impactors' HSEs.

\subsection{The entrainment of metal diapirs in the mantle}
\label{sec:metal_entrainment_mantle}

Here, we consider that the boundary between the liquid and solid mantle is defined by the fraction of silicate crystals; when this fraction exceeds $\sim 60\%$, the viscosity increases by orders of magnitude and the convective dynamics switches from turbulent to laminar \cite{Lejeune1995, Solomatov2000, Costa2005}. Thick metal ponds that accumulate at this rheological boundary are \remove{unstable}\add{subject} to Rayleigh-Taylor instabilities \cite{KaratoRamaMurthy1997}. These instabilities form large metal diapirs that are comparable in size to the metal pond they originate from \cite{KaratoRamaMurthy1997, Fleck2018, Olson2008}. We therefore consider a metal diapir of diameter $d$, which is on the same order as the impactor core diameter, settling in a solid, or partially solid mantle (see figure~\ref{fig:HSE_accretion_schematic}b). 
\add{The diapir's exact size will depend on the impactor's metal content (i.e., Fe-Ni mass fraction), which for ordinary and enstatite chondrites varies in the approximate range $f_{\rm met}\sim5-30\,$\% \protect\cite <e.g.,>{ScottKrot2003}. The diapir will thus be a fraction $(f_{\rm met} \rho_{\rm c}/\rho_{\rm FeNi})^{1/3}\sim0.3-0.5$ of the impactor's diameter.}

As in the magma ocean, these diapirs are re-entrained by convection only when the Shields number $\theta_S$ (equation~\ref{eq:shields_number}) exceeds $\theta_c=0.15\pm0.05$ \cite{Solomatov1993, Sturtz2021, Monteux2023}. However, in a high-viscosity laminar mantle, the shear stresses are now proportional to the convective speed \cite{Solomatov1993},
\begin{linenomath*}
    \begin{equation}
        \label{eq:tau_laminar}
        \tau \propto \dfrac{\mu_s\,U}{H}.
    \end{equation}
\end{linenomath*}

We expect the kinematic viscosity of a solid or partially solid mantle to be in the range $\nu_s \sim 10^{13}-10^{15}\,{\rm m}^2\,{\rm s}^{-1}$ \cite{Ita1998, Solomatov2000, Solomatov2015}, and the thermal diffusivity of the order $k_s \sim 10^{-6}\,{\rm m}^2\,{\rm s}^{-1}$ \cite{Freitas2021}. These values are more than nineteen orders of magnitude apart, such that momentum diffusivity will dominate energy transfer, and the Prandtl number $Pr=\nu_s/k_s$ can be considered infinite. 

We assume stress-free boundary conditions at the top and bottom of the mantle, given both the top layer (the liquid magma ocean) and bottom layer (the liquid outer core) have viscosities orders of magnitude smaller than the partially solid mantle.
In this limit of infinite Prandtl number, and with stress-free boundary conditions, the typical convective velocity $U$ satisfies the following scaling,
\begin{linenomath*}
    \begin{equation}
        \label{eq:convective_velocity}
        U = c\frac{k_s}{H}Ra^{2/3},
    \end{equation}
\end{linenomath*}
where $Ra=\alpha g \Delta T H^3 / k\nu$ is the Rayleigh number, $\alpha$ the thermal expansion coefficient, and $\Delta T$ the temperature difference across the mantle. In previous experimental and numerical studies, the coefficient $c$ has been found to vary in the range $\sim$\,0.05-0.2 \cite{Jarvis1982, Turcotte2002, Agrusta2020}. Combining equations~(\ref{eq:shields_number}) and~(\ref{eq:convective_velocity}), we obtain the critical diapir size that a convecting mantle can potentially sustain,
\begin{linenomath*}
    \begin{equation}
        \label{eq:critical_diapir_diameter}
        d_{\text{entr.}} = c \frac{\left(k_s\, \mu_s\, \rho_s^2 \,\alpha^2\, \Delta T^2\right)^{1/3}}{\Delta \rho\, \theta_c\, g^{1/3}}.
    \end{equation}
\end{linenomath*}

\begin{figure}[t!]
    \centering
    \includegraphics[width=\textwidth]{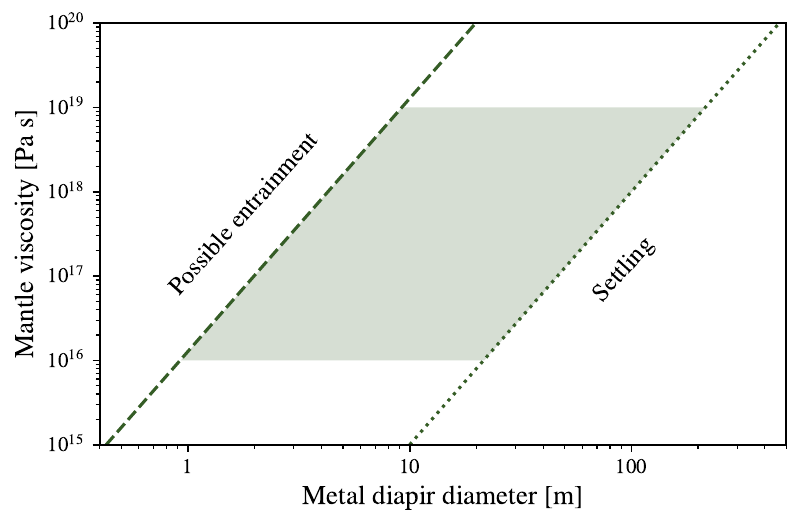}
    \caption{The maximum diameter of metal diapirs that can be entrained by mantle convection, $d_{\text{entr.}}$ (equation \ref{eq:critical_diapir_diameter}), as a function of mantle viscosity. The dashed green line corresponds to the lower estimate, using $c=0.05$, $\Delta T=500\,$K, $\theta_c=0.4$, while the dotted green line shows the higher estimate, with $c=0.02$, $\Delta T=2500\,$K, $\theta_c=0.2$. The shaded region indicates typical values expected for a partially solid mantle on the early-Earth.}
    \label{fig:CriticalDiapirRadius}
\end{figure}

This critical diapir size, shown in figure~\ref{fig:CriticalDiapirRadius}, is most sensitive to uncertainties in the mantle viscosity, which is expected to depend strongly on depth and viscosity \cite <e.g.,>{Ita1998, Solomatov2000}. Considering mantle viscosities in the range $10^{16}-10^{19}\,$Pa\,s (see Table~\ref{tab:properties} for the assumed mantle properties), this critical diapir size varies (approximately) in the range $1-100\,$m. These estimates for the critical diapir size agree with those of \citeA{KaratoRamaMurthy1997}. Larger diapirs will not be adequately supported by convective shear stresses, and will therefore sink quickly to Earth's core.

We stress that equation~(\ref{eq:critical_diapir_diameter}) is a local criterion, balancing shear stresses and the negative buoyancy of the diapir at a given depth in the mantle. When assuming a solidus temperature of about $3000\,$K at the top of the partially solid mantle \cite{Andrault2011}, temperature-dependent viscosity models \cite <adopting parameters used previously in the literature for solid mantles and mushy magma oceans; e.g.,>{RobertsZhong2006, Maurice2017} predict that a variation in temperature of about $1000\,$K throughout the partially solid mantle would cause the viscosity to vary within the range $10^{16}-10^{19}\,$Pa\,s, compatible with the range considered in figure~\ref{fig:CriticalDiapirRadius}. We therefore expect this critical diapir size, of order $1-100\,$m, to still hold when accounting for a temperature-dependent viscosity.

\subsection{Summary: HSE delivery to the mantle depends on impactor size}
\label{sec:HSE_entrainment_summary}

We have demonstrated in \S\ref{sec:post_impact_HSE_distribution} that impactors larger than $\sim\,$1\,km will readily penetrate Earth's crust, and generate significant melt (from both the target, and impactor itself). After such impacts, the impactor's metals will fragment into small drops in the newly-formed magma pond. Given a typical drop size of 1\,mm \cite{Deguen2014, Maller2024}, we demonstrated in \S\ref{sec:metal_entrainment_magma_pond} that the fraction of entrained metal would be two orders of magnitude smaller than is required to account for the observed concentration of HSEs in Earth's mantle (figure~\ref{fig:suspended_metal_MO_Earth}). These metals will therefore rapidly accumulate at the base of the magma pond, forming a layer unstable to Rayleigh-Taylor instabilities. Metal diapirs, of comparable size to the impactor, will subsequently migrate into the solid mantle. With typical size much larger than the critical diameter for convective entrainment in the solid mantle ($\sim\,$1$-$100\,m; \S\ref{sec:metal_entrainment_mantle}), these diapirs will be lost to Earth's core, leaving no HSE signature in the mantle. In contrast, evidence from terrestrial impact craters suggest that smaller impactors ($\lesssim\,$1\,km) are fragmented into millimetric pieces during impact \cite{Blau1973, Melosh2005, Folco2022}. Impactors smaller than 1\,km should therefore efficiently contribute to mantle HSE signatures thanks to efficient crustal recycling on the early Earth.

The efficient delivery of HSEs to the mantle is thus strongly dependent on impactor size. Most strikingly, in the absence of some mechanism capable of disrupting core material into very small (i.e., $\lesssim\,$$0.01\,$mm) fragments in a magma pond, large differentiated planetesimals will be unable to account for Earth's mantle HSEs. These are commonly invoked as the source of Earth's HSEs \cite <e.g,.>{Bottke2010, Brasser2016}, and comprise the majority of the total mass in an asteroid-like size distribution (see figure~\ref{fig:HSE_accretion_schematic}). This therefore has significant implications for either the source of Earth's HSEs, or estimates of total mass accretion during the late veneer. We discuss this next.

\section{HSE delivery via the entrainment of small impactors}
\label{sec:small_impactors_methods}

Motivated by the results of the previous section, in which we demonstrated that metals from only impactors smaller than $\sim\,$1\,km will be convectively entrained in Earth's mantle, we consider the feasibility of these impactors as the source of observed HSEs. In particular, we investigate the implications this would have for total mass accretion during the late veneer, and its consistency with the crucial, independent observational constraint on late accretion provided by the Moon. 

A similar scenario was initially proposed by \citeA{Schlichting2012}, who invoked a late veneer sourced by a population of small ($\sim\,$10\,m) planetesimals. \citeA{Schlichting2012} demonstrated that, by forming a collisionally damped disk, these planetesimals could effectively damp the eccentricities and inclinations of the terrestrial planets, whilst also increasing Earth's gravitational cross section relative to the Moon. This was suggested to provide a simple resolution for the high HSE ratio that is observed. A challenge for this scenario, however, is that it can only reproduce an Earth-Moon HSE ratio as high as 200, inconsistent with mantle abundances that indicate this ratio is closer to $\sim\,$2000$-$3000 \cite{DayWalker2015}.
We note that \citeA{Schlichting2012} identified the lunar crust as an additional HSE reservoir, which could prevent HSE delivery to the mantle, and thereby reduce the Earth-Moon HSE ratio. Few samples however, from both the lunar regolith and impact melt breccias, record sufficiently high HSE concentrations \cite{DayWalker2015, Day2016}, suggesting the lunar crust did not provide such an effective barrier during the late veneer.

In this section we provide two further arguments against the delivery of HSEs via small planetesimals. Specifically, we show that in order to deliver sufficient mass in small bodies from an asteroid-like size distribution, an implausibly large mass must also be delivered in larger planetesimals, which would be lost to Earth's core. If, instead, there was a dearth of large planetesimals, HSEs would be delivered predominantly to the lunar crust, rather than the mantle, in tension with observational constraints \cite{Ryder2002, DayWalker2015}.

\subsection{Mass constraints: an accretion catastrophe}
\label{sec:small_impacts_mass_constraints}

In the absence of alternative evidence supporting a (collisionally damped) disk of small planetesimals in the inner Solar System, we assume they are a subset of a larger population with a power-law size-frequency distribution (SFD),
\begin{linenomath*}
    \begin{equation}
    n(D)dD = KD^{-\alpha}dD, 
\end{equation}
\end{linenomath*}
for $D \in [D_{\rm min}, D_{\rm max}]$, where $D$ is the impactor diameter, $\alpha$ the slope of the distribution, and $K$ a constant of proportionality. The assumption that it is only impactors with $D<D_{\rm crit}$ that deliver HSEs to the mantle allows us to constrain the constant of proportionality,
\begin{linenomath*}
    \begin{equation}
        K = \left(\frac{6M_{\rm HSE, \,\oplus}}{\pi \rho_{\rm imp}}\right) \left(\int_{D_{\rm min}}^{D_{\rm crit}}{D^{3 - \alpha}dD}\right)^{-1},
    \end{equation}
\end{linenomath*}
where $\rho_{\rm imp}$ is the characteristic impactor density. The Earth will unavoidably accrete excess mass, that will make no contribution to mantle HSE signatures, from impactors in the size range $[D_{\rm crit}, D_{\rm max}]$, with the total accreted mass given by
\begin{linenomath*}
    \begin{equation}
        \label{eq:m_acc_tot}
        M_{\rm acc, tot} = \int_{D_{\rm min}}^{D_{\rm max}}{\left(\frac{\pi \rho_{\rm imp} K}{6}\right)D^{3-\alpha}dD} = M_{\rm HSE, \oplus} \left(\frac{\int_{D_{\rm min}}^{D_{\rm max}}{D^{3-\alpha}dD}}{\int_{D_{\rm min}}^{D_{\rm crit}}{D^{3-\alpha}dD}}\right).
    \end{equation} 
\end{linenomath*}
\begin{figure}[t!]
    \centering
    \includegraphics[width=1\textwidth]{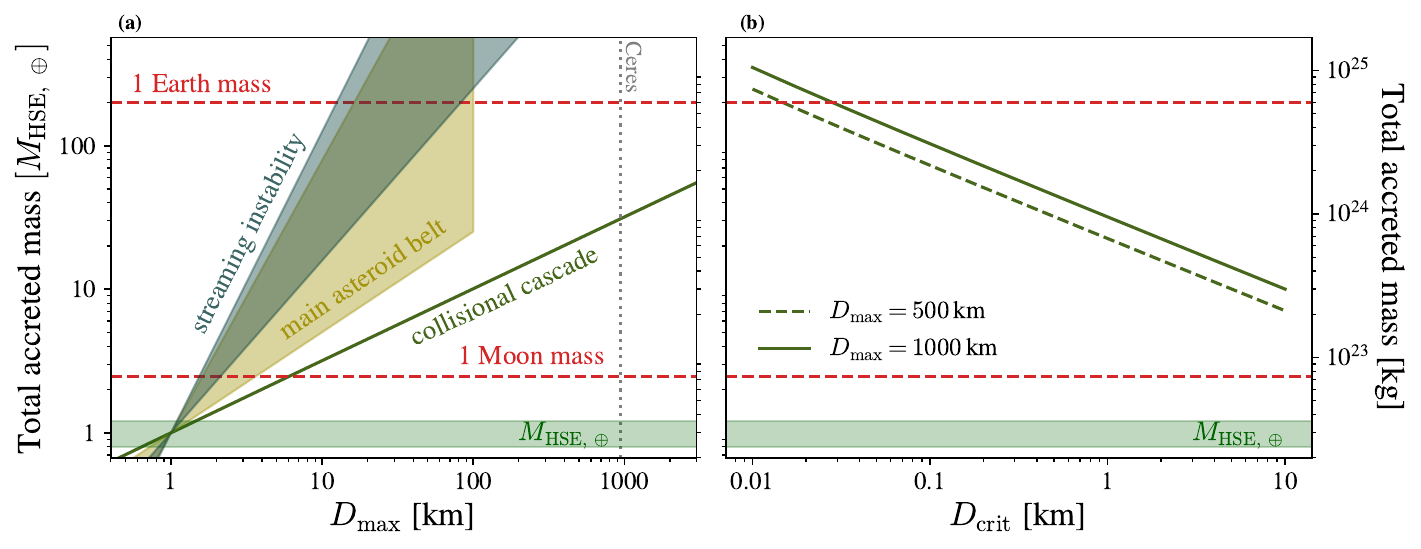}
    \caption{Total mass accretion to the Earth is calculated, assuming all observed HSEs ($M_{\rm HSE, \oplus}$) are delivered by impactors smaller than $D_{\rm crit}$, from a collisional size distribution \cite <$n(D)dD \propto D^{-7/2}dD$;>{Dohnanyi1969} with maximum impactor diameter $D_{\rm max}$. (a) We vary $D_{\rm max}$, while keeping $D_{\rm crit}=1\,$km fixed, and find total mass accretion from larger bodies quickly becomes unrealistically large. Estimates for the main asteroid belt, and streaming instability are included for reference, in blue and yellow respectively, which deliver even more mass in $D>D_{\rm crit}$ impactors. (b) Total mass accretion is calculated as a function of $D_{\rm crit}$, which is largely determined by the Earth's early-atmosphere (see~\ref{sec:appendix_melting}). Total mass accretion remains implausibly large in the presence of large ($\sim\,$500\,km) planetesimals, as are found in present-day asteroid belt, and predicted by the streaming instability.}
    \label{fig:D_max_constraints_HSE_budget}
\end{figure}

Total mass accretion during the late veneer quickly exceeds traditional estimates of $0.5\,$\% \cite{Day2007, Walker2009}, as shown in figure~\ref{fig:D_max_constraints_HSE_budget}. This is a consequence of the fact that the total mass in a collisional SFD is predominantly concentrated within the largest planetesimals (see figure~\ref{fig:HSE_accretion_schematic}). Assuming a critical diameter $D_{\rm crit}\,$$\sim\,$1\,km (see \S\ref{sec:entrainment_methods}), and a maximum impactor diameter $D_{\rm max}\,$$\sim\,$1000\,km (as observed in the present-day asteroid belt), total mass accretion is in excess of 20 Moon masses ($\sim\,$25\,\% of Earth's mass). Figure~\ref{fig:D_max_constraints_HSE_budget}\,(a) demonstrates this mass accretion catastrophe is even more problematic (with total mass accretion approaching Earth's mass) when adopting estimates for the main asteroid belt's SFD, for which $\alpha$ is in the range [2.1, 3.3] for asteroids smaller than 100\,km \cite <e.g.,>{Bottke2005, Gladman2009, Masiero2011}, or assuming a SFD inherited from the streaming instability, which constrains $\alpha$ to the (shallower) range [1.9-2.8] \cite{Johansen2015, Simon2016}.

Total mass accretion to the Earth is therefore unrealistically large during the late veneer if all observed HSEs are delivered by small $(<\,$$D_{\rm crit})$ planetesimals. 
\add{\protect As shown in figure~\ref{fig:D_max_constraints_HSE_budget}\,(b), this remains true even if the critical diameter significantly exceeds $D_{\rm crit}\sim1~$km; total mass accretion only becomes comparable to the traditional estimates of $0.5~$wt\,\% when $D_{\rm crit}$ approaches $\sim100$~km -- far in excess of the results of \S~\ref{sec:entrainment_methods}.}
The delivery of HSEs via the direct, convective entrainment of small impactors therefore precludes the existence of any larger bodies in the impactors' size distribution, in clear tension with predictions from the streaming instability \cite{Johansen2015, Simon2016}. There remains debate however regarding the origin of the largest bodies in the MAB \cite{Durda1998, Bottke2005, Morbidelli2009, Weidenschilling2011}, such that it is impossible to preclude the possibility that there was a dearth of large planetesimals during the late veneer. 

An impacting population consisting only of small planetesimals \cite <e.g.,>{Weidenschilling2011} would therefore seem the only way to avoid the mass-accretion catastrophe described in this section. As we show next, in \S\ref{sec:small_impact_cratering}, this would instead violate available constraints from the Moon's crust and mantle HSE budgets, thereby precluding the delivery of HSEs via small impactors.

\subsection{HSEs from small impactors cannot reach the lunar mantle}
\label{sec:small_impact_cratering}

Lunar mare basalts indicate the efficient mixing of HSEs into the lunar mantle from $\sim$\,$1.5\times10^{19}\,$kg of impactor material \cite{Day2007, DayWalker2015}, whilst the lunar crust records the addition of only $0.5\times10^{19}\,$ to $1.0\times10^{19}\,$kg of impactor material \cite{Ryder2002, DayWalker2015}. As previously discussed, this leaves an order of magnitude discrepancy between the observed Earth-Moon HSE ratio, and what is possible via gravitational focussing. Some process capable of removing HSEs from the lunar mantle post core-formation \remove{\protect\cite <e.g., FeS exsolution, or delayed lunar core formation;>{Rubie2016, Day2021}}\add{\protect\cite <e.g., FeS exsolution, delayed lunar core formation, or tidal driven remelting;>{Rubie2016, Day2021, Nimmo2024}} is therefore required to match this important observational constraint. A direct consequence of this requirement is that HSEs currently observed in the lunar mantle were delivered post-magma ocean crystallisation.
Small impactors must therefore be able to penetrate the thick lunar crust. We demonstrate here that this is very challenging, if not impossible, for small impactors.

\add{In contrast to the rapid solidification of Earth's mantle \protect\cite{Elkins-Tanton2008,Hamano2013}, the lunar mantle is expected to evolve slowly from its initially molten state post-formation. Samples from the lunar highlands first motivated the hypothesis that anorthite plagioclase was buoyantly segregated during magma ocean crystallisation, forming a thick anorthositic crust \cite <e.g.,>{Wood1970}. Such an insulating lid was able to efficiently regulate the rate of cooling of the magma ocean, extending the timescale over which it solidified to $\sim10-200\,$Myr \cite{ElkinsTanton2011}. Given this timescale is comparable to that of the late veneer \cite{Morbidelli2018,Brasser2020}, it is likely that many impacts will have been onto a thick crust overlaying liquid magma \cite{Jackson2023, Engels2024}. While this represents a significant deviation from classical cratering onto a solid substrate, it is sufficient for this analysis simply to determine whether small (km-scale) impactors are able to penetrate the thick lunar crust. }

The penetration depth of an impactor is estimated using crater-projectile scaling laws, with the assumption that HSEs are delivered to the mantle only when the depth of the impact crater exceeds the typical crustal thickness. We use the scaling law from \citeA{Allibert2023} \add{for the maximum crater depth ($Z_{\rm crater}$)}, which captures the transition between sub- and supersonic impacts by separating the effects of the Mach number ($M = v_{\rm imp} / U_s$), and the Froude number ($Fr = v_{\rm imp}^2/gR_{\rm imp}$),
\begin{linenomath*}
    \begin{equation}
        \frac{Z_{\rm crater}}{R_{\rm imp} Fr^{1/4}} = a \left(1 + bM^2\right)^{-c}.
    \end{equation}
\end{linenomath*}
Here $R_{\rm imp}$ is the impactor radius, $g$ the acceleration due to gravity of the Moon, and $(a,b,c)=(1.092, 0.11, 0.25)$ best-fit parameters. We assume a value for the sound speed of $U_s=4472\,{\rm ms}^{-1}$ \cite{Allibert2023}. The impact velocity is calculated following \citeA{Lissauer1988} as
\begin{linenomath*}
    \begin{equation}
        \label{eq:vimp_earth_moon}
        v_{\rm imp}^2 = v_{\rm esc, L}^2 + v_{\rm rel}^2 + 3v_{\rm kep, L}^2,
    \end{equation}
\end{linenomath*}
where $v_{\rm kep, L}$, $v_{\rm esc, L}$ are the orbital, and escape velocity of the Moon respectively, and $v_{\rm rel}$ is the impactor's relative velocity \cite <allowing for easy comparison with the characteristic velocity dispersion of a collisionally damped disk;>{Schlichting2012}. \add{Considering the maximum crater depth ensures our results are conservative: while several post-impact processes, such as jet formation during crater collapse, will limit the penetration depth of an impactor \protect\cite <e.g.,>{Landeau2021,Allibert2023,Engels2024}, no HSEs will reach the lunar mantle if the maximum crater depth is less than the crustal thickness.}
\begin{figure}[t!]
    \centering
    \includegraphics[width=\textwidth]{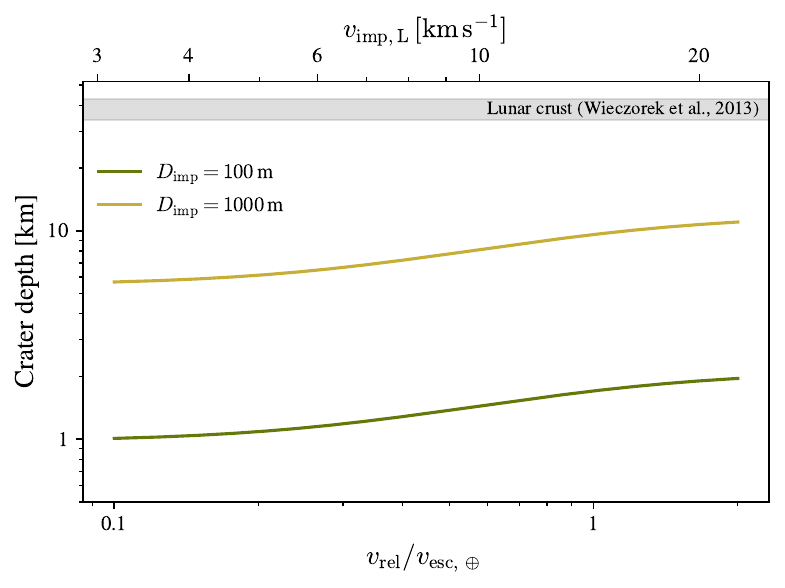}
    \caption{The impact crater depth, $Z_{\rm crater}$, is calculated as a function of relative velocity, $v_{\rm rel}$, for 100 and 1000\,m diameter impactors. Corresponding impact velocities, $v_{\rm imp, L}$, on the surface of the Moon (\remove{calcualted}\add{calculated} using equation~\ref{eq:vimp_earth_moon}) are included for reference. We use the \citeA{Allibert2023} scaling law when calculating the crater depth. In all cases, the impactors are unable to penetrate the lunar crust, which has an average depth of roughly 34\,-\,43\,km \cite{Wieczorek2013}.}
    \label{fig:small_impacts_crater_depth}
\end{figure}

\remove{Crater}\add{Maximum crater} depth, as a function of impact velocity, is shown in figure~\ref{fig:small_impacts_crater_depth}, demonstrating that small ($\lesssim\,$1\,km) impactors are unable to deliver any appreciable concentration of HSEs to the lunar mantle \add{-- in agreement with detailed hydrocode simulations of lunar impacts \protect\cite{Jackson2023}.}
These impactors will predominantly (if not uniquely) deliver HSEs to the lunar crust, which is in clear disagreement with observational constraints \cite <e.g.,>{Ryder2002, DayWalker2015}.
While it is not clear exactly when the modern lunar crust was established, we note that models support the rapid formation of a thick flotation crust as the lunar magma ocean cools \cite{ElkinsTanton2011}. Any HSEs delivered by impactors penetrating a shallow early crust would be subsequently stripped from the mantle, and would therefore not contribute to observed HSEs \cite <e.g.,>{Rubie2016, Morbidelli2018}.
It is therefore impossible to simultaneously match both the observed Earth-Moon HSE ratio, and the concentration of HSEs in the lunar mantle via the delivery of small impactors.

\subsection{Summary: the delivery of HSEs by small impactors}

We have shown in this section that for all observed HSEs to be delivered by small ($\lesssim\,$1\,km) impactors during the late veneer, an unrealistically large mass ($\lesssim\,$$M_\oplus$) would also be delivered by larger bodies, the metallic fraction of which lost to Earth's core. Thus, the only way to avoid such a mass-accretion catastrophe would be if there was a dearth of large impactors during the late veneer (i.e., a population of only sub-km impactors). We demonstrated this would also be incompatible with observational constraints, delivering HSEs predominantly (perhaps uniquely) to the lunar crust, rather than mantle. The delivery of HSEs via the direct, convective entrainment of small planetesimals is therefore unable to account for observed mantle HSEs.

\section{Summary and Discussion}
\label{sec:discussion}

In \S\ref{sec:entrainment_methods} we demonstrated that impactors larger than $\sim\,$1\,km will typically penetrate Earth's crust, and generate significant silicate melt from both the target and impactor. Using analytical scaling relations, we showed that metals from these impactors will collect at the bottom of melt pools, and that subsequent diapirism will result in HSE loss to Earth's core. However, in order to deliver sufficient mass in small impactors to account for Earth's HSEs, we demonstrated in \S\ref{sec:small_impactors_methods} that an implausibly large mass ($\lesssim\,$$M_\oplus$) would also be delivered by larger bodies in a collisional size distribution. Metals from these impactors will quickly sink to Earth's core, leaving no HSE signature in the mantle.

To avoid such a mass accretion catastrophe, our results therefore suggest that large impactors must make a significant contribution to observed mantle HSE abundances. In \S\ref{sec:resolutions_mass_accretion_catastrophe} we identify two potential resolutions to this apparent paradox, which circumvent the challenges associated with the entrainment of metal diapirs described in \S\ref{sec:entrainment_methods}. There must either exist some mechanism(s) through which HSEs, sequestered in the metallic cores of large differentiated impactors, contribute to mantle HSEs, or there was alternatively the delivery of a significant quantity of oxidised (carbonaceous chondrite-like) material during the late veneer.

In either case, there remains significant uncertainty in the efficiency of HSE delivery, and therefore total mass accretion during the late veneer. Assuming the delivery of HSEs with efficiency $f$ (whether due to the physical mechanism(s) of HSE delivery from differentiated impactors, or the mass fraction of oxidised material delivered to Earth) total mass accretion during the late veneer will be a factor $1/f$ times larger than traditional estimates. An efficiency less than 10\% would imply the late accretion of $\gtrsim\,$5\,wt\% of Earth's mass. We discuss in \S\ref{sec:disucssion_mass_constraints} the viability of such a large increase in total mass accretion in the context of four independent constraints, and comment in \S\ref{sec:methods_limitations} on the main assumptions we have made that could affect our conclusions.

\subsection{Two potential resolutions to the mass accretion catastrophe}
\label{sec:resolutions_mass_accretion_catastrophe}

\subsubsection{HSE delivery from the cores of large, differentiated impactors}

One possibility is that large, differentiated impactors successfully contribute to observed mantle HSEs. This would require the disruption of impactor core material into $\lesssim\,$$0.01\,$mm fragments, so that its metals can be convectively entrained in the impact-generated magma pond (see \S\ref{sec:metal_entrainment_magma_pond}). We reiterate, however, that this required fragment size is at least one order of magnitude smaller than is expected after an impact \cite{Deguen2014,Maller2024}, and there is presently little consensus regarding the physical mechanism(s), and efficiency of HSE delivery from the cores of differentiated planetesimals. Estimates of total mass accretion during the late veneer are therefore, currently, unconstrained.

Previous studies have focused on the accretion of lunar-sized impactors, highlighting that mass accretion may occur over a prolonged period of time due to non-merging collisions \cite{Genda2017}, or invoking the presence of a partially molten zone beneath impact-generated magma oceans \cite{KorenagaMarchi2023}. The efficiency of HSE delivery from the unavoidable collisions of smaller bodies (figure~\ref{fig:HSE_accretion_schematic}) remains, however, unclear. The cumulative contribution from these smaller bodies to Earth's mantle HSEs is therefore (currently) unconstrained and could, depending on the leftover planetesimal's SFD, significantly bias estimates of total mass accretion (see \S~\ref{sec:small_impacts_mass_constraints}). Moreover, we note that recent dynamical simulations record very few impacts of lunar-sized embryos onto the terrestrial planets post-Moon formation \cite <e.g.,>{Woo2024}.

It is possible instead that the accretion of a lunar-sized embryo is not required to account for observed HSEs. Promisingly, several recent studies \cite{Kraus2015, Li2020, Saurety2025} suggest that vapor production during large impacts has been largely underestimated during late accretion, which following the condensation and rain-out of vaporised core material, could distribute small metal droplets globally across Earth's surface. While our results (see \S\ref{sec:entrainment_methods}) demonstrate that non-vaporised metals will sink to Earth's core, the global transport of vaporised impactor core material may in principle be able to account for a significant proportion of observed HSEs \cite{Albarede2013, Kraus2015}. It is possible also that the relative dearth of lunar HSEs may arise naturally in this scenario, given the large characteristic expansion velocity of vaporised core material and low lunar escape velocity \cite{Kraus2015}.

The global distribution of Ir-rich metal nuggets in the clay layer at the K-Pg boundary is thought to be consistent with condensation from the vaporised ejecta of the ($D\sim10\,$km) Chicxulub impactor \cite{Goderis2021}, and provides some tentative observational evidence in support of the efficient delivery of HSEs via vaporisation. We note, however, that the typical size of metal droplets condensed from a vapor cloud (of crucial importance for their subsequent entrainment in a turbulent magma ocean, see \S~\ref{sec:metal_entrainment_magma_pond}) depends sensitively on both impactor diameter and impact velocity \cite{JohnsonMelosh2012}. For large planetesimals (i.e., $D\gtrsim100\,$km), the average spherule diameter is $2-3$ orders of magnitude larger than the critical droplet size $d\approx0.01\,$mm (see figure~13, \citeA{JohnsonMelosh2012}). We therefore expect that only spherules advected away from newly-formed magma ponds will effectively contribute to observed HSEs.

The vaporisation of core material therefore presents a promising avenue through which HSEs may be delivered via large, differentiated planetesimals, which will be explored in detail in future work.

\subsubsection{HSE delivery from carbonaceous chondrite-like impactors}

Alternatively, the delivery of significant quantities of oxidised carbonaceous chondrite-like material (i.e., arriving with no metal phase) would prevent the efficient loss of metals to Earth's core\add{ -- given there will be no excess density relative to the silicate melt --} and thereby help avoid a mass accretion catastrophe. The chemical composition of the late veneer is, however, a matter of longstanding debate \cite <e.g.,>{Marty2012, FischerGoddeKleine2017}. Recently however, Ruthenium isotope measurements have been argued to support the late delivery accretion of a large mass fraction ($\sim\,$60\%) of CM chondrite-like material \cite{FischerGodde2020}. \citeA{Burkhardt2021} report a slightly lower fraction ($\sim\,$30\%) on the basis of the bulk silicate Earth's Mo, supporting a substantive carbonaceous contribution to Earth's late accretion.

This picture is supported by recent studies demonstrating that the accretion of undifferentiated CC-like bodies could account for a significant fraction of Earth's Zn budget \cite{Martins2023, Martins2024}. We note further that, no matter the timing of the Moon-forming impact, the late accretion of approximately 30\% carbonaceous chondrites is recovered by dynamical simulations of solar system formation in the context of an early instability \cite{Joiret2024}, in agreement with these recent geochemical constraints. 
\begin{figure}[t!]
    \centering
    \includegraphics[width=\linewidth]{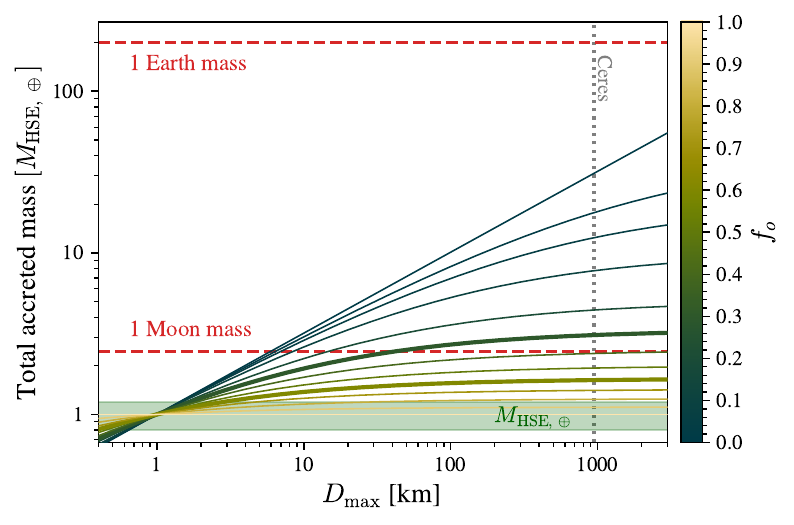}
    \caption{Total mass accretion to the Earth is calculated as a function of maximum impactor diameter, $D_{\rm max}$, assuming a collisional size distribution \cite <$n(D)dD\propto D^{-7/2}dD$;>{Dohnanyi1969}. All impactors smaller than $D_{\rm crit}=1\,$km are assumed to be entrained in Earth's mantle. Moreover, a mass fraction $f_o$ of larger planetesimals are assumed to have no metal phase, and therefore also contribute to mantle HSE signatures. This calculation, following closely the analysis in \S\ref{sec:small_impacts_mass_constraints}, is described in detail in \ref{sec:appendix_oxidised}. The bold green line corresponds to $f_o=0.6$, the mass fraction reported in \citeA{FischerGodde2020}; the bold blue line corresponds to $f_o=0.3$, as reported in \citeA{Burkhardt2021}. Total mass accretion will increase for steeper SFDs (as inferred for the main asteroid belt, and expected for the streaming instability; see figure~\ref{fig:D_max_constraints_HSE_budget}), therefore requiring a more oxidised late veneer in order to avoid a mass accretion catastrophe.}
    \label{fig:oxidising_late_veneer}
\end{figure}

Given that HSEs from the remaining fraction of reduced impactors will still be lost to Earth's core, estimates of total mass accretion remain accordingly very sensitive to the mass fraction of oxidised impactors delivered during the late veneer (see figure~\ref{fig:oxidising_late_veneer}). The late accretion of $30\,$\% carbonaceous chondrite-like material during the late veneer, as reported in \citeA{Burkhardt2021}, would require total mass accretion only $\sim\,3$ times larger than traditional estimates, which is likely consistent with independent constraints on total mass accretion to the early Earth (see \S\ref{sec:disucssion_mass_constraints}). We note, however, that while able to account for the delivery of HSEs to Earth's mantle, it remains challenging to explain the high Earth-Moon HSE ratio in this scenario, particularly given that the stochastic accretion of lunar-sized embryos post-Moon formation is not commonly observed in dynamical simulations of terrestrial planet formation \cite <e.g.,>{Woo2024}.

\subsection{How much excess mass accretion can the Earth-Moon system accommodate?}
\label{sec:disucssion_mass_constraints}

First, increased mass accretion during the late veneer would have significant implications for the geological evolution of the Hadean Earth, causing the mixing, burial, and melting of its early crust. The cumulative delivery of $\sim\,$0.15\,wt\% over 100\,Myr is able to (subject to assumptions about crustal thickness) melt all of Earth's crust \cite{Mojzsis2019}, whilst \citeA{Marchi2014} claim this bombardment would comfortably bury Earth's crust under impact melt. Mass fluxes in excess of $\sim\,$5\,wt\% would therefore be strongly in tension with the existence of zircons dating to 4.4\,Gya \cite{Valley2014}, which would not survive such an intense period of late accretion.

Second, there exist additional isotopic constraints on late accretion thanks to the isotopic similarity of the Earth, and Moon. These include the difference in $^{182}\text{W}$ between the Earth and Moon, which has been attributed to the late accretion of $\sim\,$0.7\,wt\,\% of chondritic material \cite{Touboul2015, Kruijer2015}. \citeA{Jacobson2014} also infer an upper limit for late accretion of about 0.01\,$M_\oplus$, based on the O and Ti isotopic similarity of the Earth and Moon. Given the approximate consensus between these separate isotopic constraints, it appears likely that mass accretion significantly in excess of 0.01\,$M_\oplus$ would require increasingly fine-tuned assumptions about the isotopic compositions of late accreted planetesimals.

Third, various dynamical simulations find a negative correlation between the late veneer mass, and the timing of the Moon-forming impact \cite <e.g.,>{Jacobson2014, Woo2024}. This has a simple qualitative explanation; there will be fewer available planetesimals to source a late veneer at later times (a consequence of collisions with planets, ejection onto hyperbolic orbits, and catastrophic collisions with other planetesimals). Whilst these results are clearly model-specific, both \citeA{Jacobson2014} and \citeA{Woo2024} suggest the late accretion of $\sim\,$5\,wt\% would require Moon-formation earlier than $\sim\,$20\,Myr after the condensation of the first solids in the solar system. This is highly inconsistent with the date of the Moon forming impact, as recorded by various radioactive chronometers \cite <e.g.,>{Kleine2005, Touboul2007}.

Fourth, impact-generated satellites (e.g, the Moon) are very sensitive to ongoing accretion \cite{PahlevanMorbidelli2015}. Impacts onto the Earth are typically preceded by many ($\sim\,$$10^4$) collisionless encounters, which efficiently transfer angular momentum to the Moon. The late accretion of $\sim\,$5\,wt\% would therefore likely drive significantly increased orbital excitation of the Moon, and most likely its dynamical loss \cite{PahlevanMorbidelli2015}. We note, requiring earlier Moon-formation to accommodate the late delivery of $\sim\,$5\,wt\% seems particularly incompatible with the long-term stability of the Moon's orbit.

It is therefore highly improbable that the late delivery of HSEs to Earth's mantle was particularly inefficient (i.e., less than $\sim\,$10\%), given this would imply a large mass flux inconsistent with several independent constraints on total mass accretion to the Hadean Earth.

\subsection{Model limitations}
\label{sec:methods_limitations}

There are a number of caveats to our conclusions, concerning both the entrainment of metals in the mantle, and the implications for Earth's accretion history. We summarise the most important here.

Estimates of excess mass accretion during the late veneer, leaving no mantle HSE signature, are sensitive to the critical impactor diameter, $D_{\rm crit}$. As motivated in \S\ref{sec:post_impact_HSE_distribution}, $D_{\rm crit}$ roughly separates the efficient and inefficient delivery of HSEs to Earth's mantle (from smaller and larger bodies respectively). We identify criteria for melting, and penetrating Earth's crust as crucial in determining $D_{\rm crit}$. Our analysis does not, however, account for all relevant physical processes, and it is therefore possible that we over-, or underestimate the true value of $D_{\rm crit}$. This uncertainty will accordingly decrease (increase) excess mass accretion during the late veneer. As shown in Figure~\ref{fig:D_max_constraints_HSE_budget}, however, an order of magnitude increase in $D_{\rm crit}$ would still imply total mass accretion incompatible with several independent constraints (\S\ref{sec:disucssion_mass_constraints}). Unless this critical diameter is much larger than $\sim\,$10\,km, this uncertainty is unlikely to alter our main conclusion: small planetesimals are unable to account for observed HSEs, and there is correspondingly significant uncertainty in total mass accretion during the late veneer.

The entrainment of metals in a magma pond is very sensitive to the radius of the metal drops, an empirically determined efficiency factor ($\epsilon$), and the heat flux at the surface of the magma pond ($F$). The radius of metal drops following a planetary impact is however poorly known, and estimates for this empirical efficiency factor $\epsilon$ differ by nearly a factor of 5 \cite{Solomatov1993, LavorelLeBars2009}. Accordingly, there is uncertainty in the equilibrium mass of suspended metal, as is evident in figure~\ref{fig:suspended_metal_MO_Earth}. Even when taking end-member values for each parameter ($\epsilon$ and $F$), it is not possible to account for Earth's HSEs. For our conclusions to change significantly, alternative fragmentation mechanisms must exist that can generate metal drops smaller than $0.01\,$mm. We reiterate, however, that currently this remains at least one order of magnitude below best-estimates \cite{Deguen2014, Maller2024}, and any mechanism capable of achieving this remains elusive.

As discussed in \S\ref{sec:metal_entrainment_magma_pond}, the condition used to determine the entrainment of metal drops in a magma pond ($\theta_S>\theta_c$) has only been demonstrated for solid particles \cite{Solomatov1993, Sturtz2021}. This has not however been tested for liquid drops, and there is therefore uncertainty in our results regarding the critical drop size for entrainment. Liquid drops can in principle merge with the bottom layer, and it is therefore possible that surface tension may in fact decrease the efficiency of this re-entrainment processes. We therefore expect that our conclusions still hold despite the lack of experiment on the sedimentation of dilute liquid drops in magma oceans.
\add{\protect We additionally expect our results to be robust to uncertainty in the power law size distribution of drops formed in an impact-generated magma ocean (see~\ref{sec:appendix_drop_distribution}), provided the total metal mass is dominated by the largest drops as is reported in \citeA{Riviere2022}.}

\add{\protect We implicitly assume throughout our analysis in \S~\ref{sec:metal_entrainment_magma_pond} that the settling timescale of metal droplets is much shorter than the solidification timescale of the impact-generated magma pond. If, however, the magma pond solidifies sufficiently quickly, metal droplets may be frozen in place before they accumulate into the gravitationally unstable metal layer at its base. We expect our analysis to be robust to this assumption, particularly for the largest impacts that dominate total mass accretion (figure~\ref{fig:HSE_accretion_schematic}), with the solidification timescale of (local) magma ponds \cite <of order $100~$yrs;>{Rubie2003,ReeseSolomatov2006} several orders of magnitude greater than the settling timescale of metal droplets in even deep magma oceans \cite <of order weeks--months;>{Rubie2003}.}

Finally, the condition used to determine the entrainment of metal diapirs in a solid, or partially solid, mantle neglects the possible deformation of the diapir by the convective flow and the multiphase dynamics of mushy, solidifying magma oceans \cite{Ballmer2017, Maurice2017, Morison2019, Labrosse2024, Boukare2025}. Quantifying these effects requires further investigation using numerical simulations of convecting mushy magma oceans, which is beyond the scope of this paper.

\section{Conclusions}

As recorded by the elevated concentrations of HSEs in the mantle, the late veneer is thought to deliver at least $\sim\,$0.5\,wt\% (of Earth's mass) of chondritic material to the Earth, and $\sim\,$0.02\,wt\% to the Moon \cite{Day2016}. A consequence of the negative buoyancy of metals in both fully liquid magma ponds and the solid mantle, however, is that the entrainment of HSEs in Earth's mantle is much more challenging than previously thought. We demonstrate in this study that there exists a critical diameter ($\sim\,$1\,km), separating the efficient and inefficient delivery of HSEs to the mantle, with the direct entrainment of HSEs possible only for smaller impactors via the tectonic recycling of the crust. However, despite the potentially efficient delivery of metals to Earth's mantle by small impactors, we demonstrate that they cannot be the dominant source of Earth's HSEs. In order to deliver sufficient mass in sub-km bodies, we show there would be an implausibly large mass ($\lesssim\,$1\,$M_\oplus$) delivered by larger ($D>\,$1\,km) bodies in a collisional size distribution, with their HSEs lost to Earth's core. There is therefore, currently, a contradiction between the observed concentrations of HSEs in the mantle, the geodynamics of metal entrainment, and estimates of total mass accretion during the late veneer.

To avoid such a mass accretion catastrophe, we suggest impactors larger than 1\,km must make significant contributions to mantle HSE signatures, and identify two potential resolutions to this apparent paradox. We show that it would be possible to suspend sufficient HSEs in an impact-generated magma pond, if there exists mechanisms able to disrupt impactor core material into $\lesssim\,$$0.01$\,mm fragments during large planetary impacts. This is, however, at least one order of magnitude smaller than current estimates for the size of metal drops after impacts.  Alternatively, the delivery of a significant mass fraction of oxidised material during the late veneer would prevent the loss of metals to Earth's core, thereby avoiding such a mass accretion catastrophe. In both scenarios, total mass accretion during the late veneer remains unconstrained, due to uncertainty in the efficiency of physical mechanisms capable of disrupting impactor core material into sufficiently small fragments, and the mass fraction of oxidised material delivered to Earth during the late veneer.



\appendix
\section{The condition for melting by impact}
\label{sec:appendix_melting}

\subsection{Quantification of impact-induced melting}

The fate of impactor material is dependent on the peak pressure experienced during the compression stage of crater formation, as this will directly determine the density, and temperature of impactor material. Regions of impactor material will melt (vaporise) if they experiences peak shock pressures in excess of the material's incipient melting (vaporisation) shock pressure. 

The distribution of shock pressure within projectiles is however highly complex, and so detailed hydrocode simulations are necessary to accurately determine the fate of impactor material \cite <e.g.,>{PierazzoMelosh2000, PotterCollins2013}. Here, we use the results from iSALE simulations in \citeA{PotterCollins2013} to estimate the fraction of impactor material that does not melt, 
\begin{linenomath*}
    \begin{equation}
        \label{eq:PotterCollins_meltfrac}
        f = 1 - \cos^{1.3}{\left(\frac{\pi}{2}\frac{P_{\rm melt}}{P_{\rm max}\sin{\theta}}\right)},
    \end{equation}
\end{linenomath*}
where $P_{\rm melt}$ is the critical shock pressure for incipient melting \cite <assumed to be the ANEOS-derived value of 106\,GPa for dunite projectiles;>{Wunnemann2008}, $P_{\rm max}$ the peak shock pressure experienced anywhere in the impactor, and $\theta$ the impact angle. Note, the choice of dunite as a proxy for asteroidal composition ensures that our estimated critical impactor size for melting, $D_{\rm crit}$ provides an upper limit, given the critical pressure for melting is significantly lower for other rock types \cite{Wunnemann2008}.
\begin{figure}[t]
    \centering
    \includegraphics[width=0.9\textwidth]{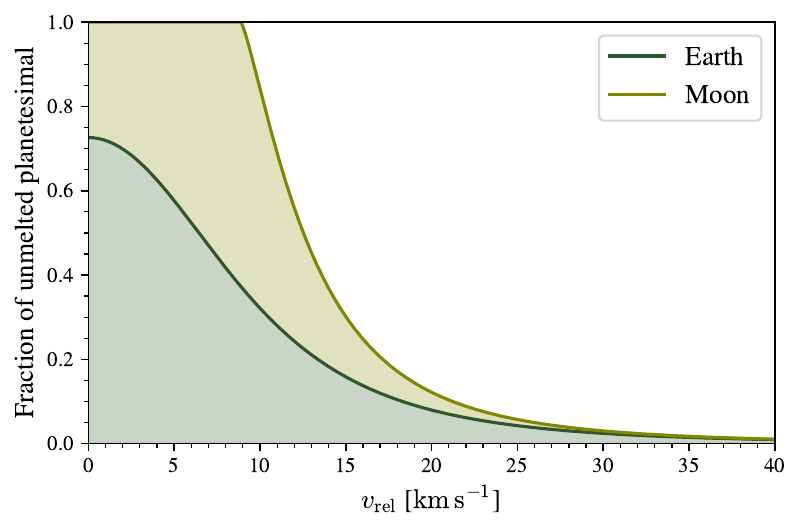}
    \caption{The fraction of unmelted impactor material is plotted as a function of relative velocity, $v_{\rm rel}$, for both the Earth and Moon. This is calculated using the parameterisation for asteroid survivability from \citeA{PotterCollins2013}, assuming a critical shock pressure for melting of 106\,GPa \cite{Wunnemann2008}. The material constants $(C_i, S_i) = (5.430\,{\rm km\,s}^{-1}, 1.34)$ and $(C_t, S_t) = (3.816\,{\rm km\,s}^{-1}, 1.28)$ are chosen for a dunite impactor and granite target \cite <Table 1;>{PotterCollins2013}. Significant melting will occur onto both the Earth and Moon for a typical relative velocity of $15\,{\rm km}\,{\rm s}^{-1}$.}
    \label{fig:impactor_melting_vrel}
\end{figure}

The peak shock pressure, $P_{\rm max}$, is estimated using the planar shock approximation, which is relatively accurate for materials with a linear shock-particle velocity relationship  \cite <i.e., $U = C + Su$, where $C$ and $S$ are empirically determined parameters describing target, or impactor material;>{Melosh1989}. The peak pressure is given by the Hugoniot equation 
\begin{linenomath*}
    \begin{equation}
        P_{\rm max} = \rho_{0i} u_i \left(C_i + S_iu_i\right),
    \end{equation}
\end{linenomath*}
where $\rho_{0i}$ is the uncompressed impactor density, and $u_i$ the impactor particle velocity. This peak pressure increases with impact velocity, $v_{\rm imp}$, through its dependence on the particle velocity, which is given by
\begin{linenomath*}
    \begin{equation}
        u_i = \frac{-B + \sqrt{B^2 - 4AC}}{2A},
    \end{equation}
\end{linenomath*}
where
\begin{linenomath*}
    \begin{align}
        A &= \rho_{0i}S_i - \rho_{0t} S_t,\\
        B &= \rho_{0i}C_i + \rho_{0t}C_t + 2 \rho_{0t} S_t v_{\rm imp},\\
        C &= -\rho_{0t}v_{\rm imp}\left(C_t + S_t v_{\rm imp}\right),
    \end{align}
\end{linenomath*}
with the subscripts $i,t$ referring to impactor and target material respectively.

The fraction of unmelted impactor material is plotted as a function of relative velocity, $v_{\rm rel}$, in Figure~\ref{fig:impactor_melting_vrel}. The corresponding impact velocities on the Moon and Earth are given by
\begin{linenomath*}
    \begin{align}
        v_{\rm imp,\,\oplus}^2 &= {v_{\rm esc,\,\oplus}^2 + v_{\rm rel}^2},\\
        v_{\rm imp,\,L}^2 &= {v_{\rm esc,\,L}^2 + 3v_{\rm kep,\,L}^2 + v_{\rm rel}^2}
    \end{align}
\end{linenomath*}
respectively \cite <e.g.,>{Lissauer1988}, allowing for easy comparison between impacts on the Earth and Moon.

\subsection{The atmospheric entry of planetesimals}

Impactors entering Earth's atmosphere are subject to large ram pressures, often leading to the significant deceleration and fragmentation of small bodies. The size and velocity of impactors arriving at the top of Earth's atmosphere may therefore be considerably different to those reaching the surface. Here we determine the minimum impactor diameter for melting, assuming a critical impact velocity of $15\,{\rm km}\,{\rm s}^{-1}$ (as determined in the previous section).

The trajectory of an impactor through the atmosphere is first described by a set of four coupled differential equations \cite <e.g.,>{PasseyMelosh1980}, before fragmentation occurs when the ram pressure exceeds the bodies tensile strength. Detailed modelling of fragmentation is however challenging, given the complex physical processes involved. We thus use the simple 1D model from \citeA{Chyba1993}, which is able to reproduce the observed energy deposition of Tunguska-like impactors, describing the deformation (or ``pancaking'') of impactor material into a cylindrical shape.

We assume the impactor will fragment into a number of small pieces (in an airburst-like event) when the radius of the impactor reaches 6 times its initial value \cite <following>{Chyba1993}. To determine the critical diameter for melting, we numerically calculate the trajectory of a stony projectile, varying its initial diameter until it is able to reach the surface intact, with velocity greater than 15\,km\,s$^{-1}$. We assume an isothermal atmospheric profile, with a scale height of 7\,km, and vary the atmospheric surface density. 
\begin{figure}[t]
    \centering
    \includegraphics[width=0.9\textwidth]{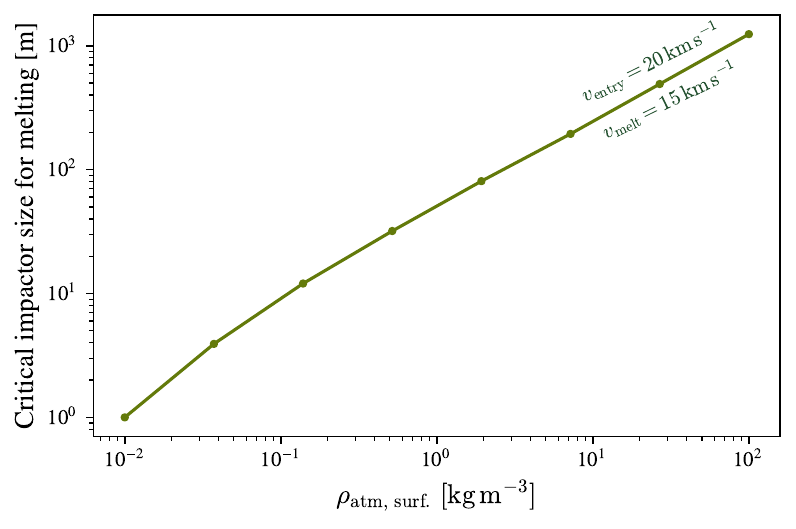}
    \caption{The critical impactor size for melting is calculated as a function of atmospheric surface density, $\rho_{\rm atm, surf.}$, using the deformation model from \citeA{Chyba1993}, assuming an isothermal atmospheric profile with a constant scale height of 8\,km. A constant entry velocity of 20\,km\,s$^{-1}$ is used, with significant melting expected for impact velocities in excess of 15\,km\,s$^{-1}$ (see figure~\ref{fig:impactor_melting_vrel}).}
    \label{fig:critical_impactor_size_melting}
\end{figure}

The results of this calculation are shown in figure~\ref{fig:critical_impactor_size_melting}, in which we see for a 1\,bar atmosphere, the critical radius for melting, $R_m$, is roughly 50\,m. This increases to roughly 1000\,m for a 100\,bar atmosphere, reflecting the increased ram pressure experienced by the impactor.

\section{\add{Metal drop size distribution in impact-generated magma pond}}
\label{sec:appendix_drop_distribution}

\add{Recent results on the size distribution of bubbles in a turbulent flow \protect\cite<e.g.,>{Riviere2022} demonstrate that the drop size distribution follows a power law. Below the Hinze scale, at which surface tension equilibrates local pressure fluctuations -- expected to determine the mean drop size in an impact-generated magma ocean \cite{Deguen2014,WacheulLeBars2018} -- drop size scales as $n(d)\propto d^{-3/2}$, where $d$ is the drop diameter \cite{Riviere2022}. For such a power law, the mass in drops of size $d$ varies as $m(d) \propto d^{3/2}$, meaning that the total metal mass is dominated by the largest drops present in the magma ocean. Here we demonstrate that, while small drops of size $\sim\,0.01\,{\rm mm}$ might be present in a magma ocean, they will carry insufficient mass to account for Earth's HSEs.}

\begin{figure}[t]
    \centering
    \includegraphics[width=0.9\textwidth]{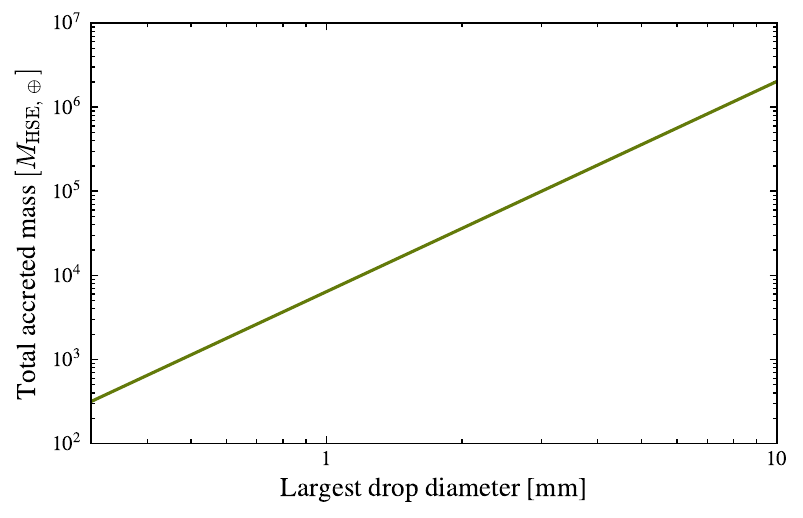}
    \add{\protect\caption{The required late veneer mass is calculated as a function of the largest drop diameter, assuming that HSEs were delivered in drops of diameter smaller than $0.03\,$mm, which could be entrained in significant proportions in the impact-generated magma. The drop size distribution is assumed to follow a $d^{-3/2}$ power-law, where $d$ is the drops diameter \cite{Riviere2022}. With the largest drop size expected to be $\sim\,1\,$mm \cite{Deguen2014,WacheulLeBars2018,Maller2024}, this would require an unreasonably large late accreted mass in excess of $6000\,M_{\rm HSE, \oplus}$.}}
    \label{fig:supplementary_fig}
\end{figure}

\add{We consider a distribution of metal drops with diameters in the range $[d_{\rm min}, d_{\rm max}]$, where $d_{\rm max} \gg d_{\rm min}$ is the size of the largest drops in the impact-generated magma. Figure~\protect\ref{fig:suspended_metal_MO_Earth} demonstrates that only metal drops smaller than $d_{\rm ent.} \sim 0.01\,$mm can potentially be entrained with sufficiently high volume ratio to explain Earth's HSEs. The total mass in suspended drops ($d\leq d_{\rm ent.}$) is thus given by
\begin{equation}
    \label{eq:suspended_small_drops}
    \dfrac{M(d\leq d_{\rm ent.})}{M_{\rm HSE, \oplus}} = \dfrac{d_{\rm ent.}^{5/2} - d_{\rm min}^{5/2}}{d_{\rm max}^{5/2} - d_{\rm min}^{5/2}} \sim \left(\dfrac{d_{\rm ent.}}{d_{\rm max}}\right)^{5/2}.
\end{equation}
From equation~\ref{eq:suspended_small_drops} it is clear that the required late veneer mass must significantly exceed $M_{\rm HSE, \oplus}$ when $d_{\rm ent.}<d_{\rm max}$. This is illustrated in figure~\ref{fig:supplementary_fig}, which shows the required late veneer mass as a function of $d_{\rm max}$, assuming that HSEs were delivered by metal drops smaller than $d_{\rm ent.}=0.01\,{\rm mm}$, $0.03\,{\rm mm}$, and $0.1\,{\rm mm}$. Assuming the largest drops have a diameter $\sim 1\,{\rm mm}$ \cite{Deguen2014,WacheulLeBars2018,Maller2024}, this would require an implausibly large late accreted mass of $6000\,M_{\rm HSE, \oplus}~(> M_\oplus)$. Thus, while there will be a distribution of droplets produced following a large impact, extending down to small sizes, these small drops will collectively carry insufficient mass to account for Earth's HSEs.}

\section{An oxidized late veneer}
\label{sec:appendix_oxidised}

The accretion of oxidised carbonaceous-chondrite material during the late veneer is one potential way to avoid the loss of metal, and its HSEs, to Earth's core (as described in \S\ref{sec:entrainment_methods}). This material, arriving with no (or very small amounts) metal phase present will efficiently contribute its HSEs to the mantle, by virtue of its much smaller density contrast with the surrounding silicates. 

Here, we present a simple calculation assuming some constant (i.e., size-independent) fraction $f_o$ of material arrives with no metal phase, with its HSEs accordingly distributed throughout its silicate components, and therefore contributes to observed HSE signatures. Following our approach in \S\ref{sec:small_impacts_mass_constraints}, we assume a collisional size distribution \cite <$n(D)dD = K D^{-7/2}dD$>{Dohnanyi1969}, and assume all impactors smaller than $D_{\rm crit}=1\,$km are also entrained in Earth's mantle. The SFDs constant of proportionality, $K$, is now given by
\begin{linenomath*}
    \begin{equation}
        K = \left(\frac{6M_{\rm HSE,\oplus}}{\pi \rho_{\rm imp}}\right) \left[\int_{D_{\rm min}}^{D_{\rm crit}}{D^{3-\alpha}dD} + f_o\int_{D_{\rm crit}}^{D_{\rm max}}{D^{3-\alpha}dD}\right]^{-1}.
    \end{equation}
\end{linenomath*}
Total mass accretion is therefore
\begin{linenomath*}
    \begin{equation}
        M_{\rm acc, tot} = M_{\rm HSE,\oplus} \Bigg[\frac{\int_{D_{\rm min}}^{D_{\rm max}}{D^{3-\alpha}dD}}{\int_{D_{\rm min}}^{D_{\rm crit}}{D^{3-\alpha}dD} + f_o\int_{D_{\rm crit}}^{D_{\rm max}}{D^{3-\alpha}dD}}\Bigg],
    \end{equation}
\end{linenomath*}
which reverts exactly to equation~\ref{eq:m_acc_tot} in the limit $f_o\rightarrow0$, as expected. Total mass accretion, as a function of the SFDs maximum impactor diameter $D_{\rm max}$, is shown in Figure~\ref{fig:oxidising_late_veneer}. Increasing the mass fraction of oxidised material delivered (i.e., increasing $f_o$) effectively helps avoid the mass accretion catastrophe. Total mass accretion will remain less than 1 Moon mass provided $f_o\gtrsim0.4$, which is smaller than the mass fraction reported in \citeA{FischerGodde2020}. This is only $\sim\,$3 times larger than traditional estimates of total mass accretion during the late veneer, and is likely consistent with independent constraints on total mass accretion to the Hadean Earth (see \S\ref{sec:disucssion_mass_constraints}).

It is a significant simplification to assume the oxidation state of planetesimals is size-independent, and it is possible instead that this fraction could decrease significantly for larger bodies able to form a metallic core. This would significantly increase total mass accretion, and we would quickly return to the mass accretion catastrophe described in \S\ref{sec:small_impactors_methods}, given that total mass is concentrated within the largest bodies in a collisional size distribution. 

Evidence from within the Solar System is, however, not totally conclusive. Whilst some oxidised bodies formed with metallic, iron-rich cores, as is evidenced by both iron meteorites \cite{Grewal2024} and the water-rich Jovian satellite Ganymede \cite{Schubert1996}, Ceres (the largest body in the asteroid belt) demonstrates that others did not \cite{Thomas2005}. We therefore, in the interests of simplicity, assume this fraction is size-independent, and note this has the potential to significantly bias the results shown in figure~\ref{fig:oxidising_late_veneer}.

%



%
%

\section*{Open Research Section}

\remove{No new data were generated for this work. Scripts used to generate the figures presented in this work have been are publicly available at \protect\url{https://github.com/richard17a/HSE_mass_accretion}. This repository will be archived in Zenodo upon acceptance.}
\add{\protect No new data were generated for this work. Scripts used to generate the figures presented in this work have been archived in Zenodo: \url{https://zenodo.org/records/18963669} \cite{Anslow2026_zenodo}, and are also available at \url{https://github.com/richard17a/HSE_mass_accretion}.}

\acknowledgments
We thank Helen Williams, Ingrid Blanchard, and Julien Siebert for useful discussions. R.J.A. acknowledges the Science and Technology Facilities Council (STFC) for a PhD studentship. A.B. acknowledges the support of a Royal Society University Research Fellowship, URF/R1/211421. This work was supported by the Programme National de Planétologie (PNP) of CNRS-INSU, co-funded by CNES. 

\add{\protect\section*{Conflict of Interest Statement}}
\add{The authors have no conflicts of interest to disclose.}

%
%

\bibliography{example}

@article{Solomatov1993,
  title={Entrainment from a bed of particles by thermal convection},
  author={Solomatov, Viatcheslav S and Olson, Peter and Stevenson, David J},
  journal={Earth and planetary science letters},
  volume={120},
  number={3-4},
  pages={387--393},
  year={1993},
  publisher={Elsevier}
}

@ARTICLE{Riviere2022,
       author = {{Rivi{\`e}re}, Ali{\'e}nor and {Ruth}, Daniel J. and {Mostert}, Wouter and {Deike}, Luc and {Perrard}, St{\'e}phane},
        title = "{Capillary driven fragmentation of large gas bubbles in turbulence}",
      journal = {Physical Review Fluids},
         year = 2022,
        month = aug,
       volume = {7},
       number = {8},
          eid = {083602},
        pages = {083602},
          doi = {10.1103/PhysRevFluids.7.083602},
archivePrefix = {arXiv},
       eprint = {2112.06480},
 primaryClass = {physics.flu-dyn},
       adsurl = {https://ui.adsabs.harvard.edu/abs/2022PhRvF...7h3602R}
}

@ARTICLE{DayPaquet2021,
       author = {{Day}, James M.~D. and {Paquet}, Marine and {Timothy Jull}, A.~J.},
        title = "{Temporally limited late accretion after core formation in the Moon}",
      journal = {Meteoritics \& Planetary Science},
         year = 2021,
        month = apr,
       volume = {56},
       number = {4},
        pages = {683-699},
          doi = {10.1111/maps.13646},
       adsurl = {https://ui.adsabs.harvard.edu/abs/2021M&PS...56..683D}
}

@ARTICLE{Steenstra2020,
       author = {{Steenstra}, E.~S. and {Berndt}, J. and {Klemme}, S. and {Snape}, J.~F. and {Bullock}, E.~S. and {van Westrenen}, W.},
        title = "{The Fate of Sulfur and Chalcophile Elements During Crystallization of the Lunar Magma Ocean}",
      journal = {Journal of Geophysical Research (Planets)},
         year = 2020,
        month = nov,
       volume = {125},
       number = {11},
          eid = {e06328},
        pages = {e06328},
          doi = {10.1029/2019JE006328},
       adsurl = {https://ui.adsabs.harvard.edu/abs/2020JGRE..12506328S}
}

@article{Blanchard2025,
  title = {Earth’s deep magma ocean never reached sulfide saturation},
  author = {I. Blanchard and J. Siebert and E. Kubik and A. Minchenkova and L. Calvo and N. Wehr},
  journal = {Geochemical Perspectives Letters},
  year = {2025},
  volume = {34},
  pages = {6--10},
  doi = {https://doi.org/10.7185/geochemlet.2506},
  url = {https://www.geochemicalperspectivesletters.org/article2506}
}

@article{Solomatov1993b,
  title={Suspension in convective layers and style of differentiation of a terrestrial magma ocean},
  author={Solomatov, Viatcheslav S and Stevenson, David J},
  journal={Journal of Geophysical Research: Planets},
  volume={98},
  number={E3},
  pages={5375--5390},
  year={1993},
  publisher={Wiley Online Library}
}

@article{Sturtz2021,
  title={The fate of particles in a volumetrically heated convective fluid at high Prandtl number},
  author={Sturtz, Cyril and Kaminski, {\'E}douard and Limare, Angela and Tait, Stephen},
  journal={Journal of Fluid Mechanics},
  volume={929},
  pages={A28},
  year={2021},
  publisher={Cambridge University Press}
}

@ARTICLE{Monteux2023,
       author = {{Monteux}, J. and {Qaddah}, B. and {Andrault}, D.},
        title = "{Conditions for Segregation of a Crystal-Rich Layer Within a Convective Magma Ocean}",
      journal = {Journal of Geophysical Research (Planets)},
         year = 2023,
        month = may,
       volume = {128},
       number = {5},
          doi = {10.1029/2023JE007805},
       adsurl = {https://ui.adsabs.harvard.edu/abs/2023JGRE..12807805M}
}

@article{Solomatov2000,
  title={Fluid dynamics of a terrestrial magma ocean},
  author={Solomatov, Viatcheslav S},
  journal={Origin of the Earth and Moon},
  volume={1},
  pages={323--338},
  year={2000},
  publisher={Tucson}
}

@article{Ita1998,
  title={Diffusion in MgO at high pressure: Implications for lower mantle rheology},
  author={Ita, Joel and Cohen, Ronald E},
  journal={Geophysical research letters},
  volume={25},
  number={7},
  pages={1095--1098},
  year={1998},
  publisher={Wiley Online Library}
}

@article{Solomatov2015,
  title={Magma oceans and primordial mantle differentiation},
  author={Solomatov, V},
  journal={Treatise on geophysics, Earth Formation and Evolution},
  volume={9},
  pages={91--119},
  year={2015}
}

@article{Shraiman1990,
  title={Heat transport in high-Rayleigh-number convection},
  author={Shraiman, Boris I and Siggia, Eric D},
  journal={Physical Review A},
  volume={42},
  number={6},
  pages={3650},
  year={1990},
  publisher={APS}
}

@article{Agrusta2020,
  title={Mantle convection interacting with magma oceans},
  author={Agrusta, Roberto and Morison, A and Labrosse, St{\'e}phane and Deguen, R and Alboussi{\`e}re, T and Tackley, PJ and Dubuffet, F},
  journal={Geophysical Journal International},
  volume={220},
  number={3},
  pages={1878--1892},
  year={2020},
  publisher={Oxford University Press}
}

@book{Turcotte2002,
  title={Geodynamics},
  author={Turcotte, Donald L and Schubert, Gerald},
  year={2002},
  publisher={Cambridge university press}
}

@article{Jarvis1982,
  title={Mantle convection as a boundary layer phenomenon},
  author={Jarvis, Gary T and Peltier, WR},
  journal={Geophysical Journal International},
  volume={68},
  number={2},
  pages={389--427},
  year={1982},
  publisher={Blackwell Publishing Ltd Oxford, UK}
}

@article{Pierazzo1997,
  title={A reevaluation of impact melt production},
  author={Pierazzo, E and Vickery, AM and Melosh, HJ},
  journal={Icarus},
  volume={127},
  number={2},
  pages={408--423},
  year={1997},
  publisher={Elsevier}
}

@article{Chyba1993,
  title={The 1908 Tunguska explosion: atmospheric disruption of a stony asteroid},
  author={Chyba, Christopher F and Thomas, Paul J and Zahnle, Kevin J},
  journal={Nature},
  volume={361},
  number={6407},
  pages={40--44},
  year={1993},
  publisher={Nature Publishing Group UK London}
}

@article{Melosh2005,
  title={Meteor Crater formed by low-velocity impact},
  author={Melosh, HJ and Collins, GS},
  journal={Nature},
  volume={434},
  number={7030},
  pages={157--157},
  year={2005},
  publisher={Nature Publishing Group UK London}
}

@article{Svetsov1995,
  title={Disintegration of large meteoroids in Earth's atmosphere: Theoretical models},
  author={Svetsov, VV and Nemtchinov, IV and Teterev, AV},
  journal={Icarus},
  volume={116},
  number={1},
  pages={131--153},
  year={1995},
  publisher={Elsevier}
}

@article{Blau1973,
  title={Investigation of the Canyon Diablo metallic spheroids and their relationship to the breakup of the Canyon Diablo meteorite},
  author={Blau, Peter J and Axon, Howard J and Goldstein, Joseph I},
  journal={Journal of Geophysical Research},
  volume={78},
  number={2},
  pages={363--374},
  year={1973},
  publisher={Wiley Online Library}
}

@article{Folco2022,
  title={Microscopic impactor debris at Kamil Crater (Egypt): The origin of the Fe-Ni oxide spherules},
  author={Folco, Luigi and Carone, L and D'Orazio, Massimo and Cordier, Carole and Suttle, Martin D and van Ginneken, Matthias and Masotta, M},
  journal={Geochimica et Cosmochimica Acta},
  volume={335},
  pages={297--322},
  year={2022},
  publisher={Elsevier}
}

@ARTICLE{PasseyMelosh1980,
       author = {{Passey}, Q.~R. and {Melosh}, H.~J.},
        title = "{Effects of atmospheric breakup on crater field formation}",
      journal = {Icarus},
         year = 1980,
        month = may,
       volume = {42},
       number = {2},
        pages = {211-233},
          doi = {10.1016/0019-1035(80)90072-X},
       adsurl = {https://ui.adsabs.harvard.edu/abs/1980Icar...42..211P}
}

@ARTICLE{Morbidelli2018,
       author = {{Morbidelli}, A. and {Nesvorny}, D. and {Laurenz}, V. and {Marchi}, S. and {Rubie}, D.~C. and {Elkins-Tanton}, L. and {Wieczorek}, M. and {Jacobson}, S.},
        title = "{The timeline of the lunar bombardment: Revisited}",
      journal = {Icarus},
         year = 2018,
        month = may,
       volume = {305},
        pages = {262-276},
          doi = {10.1016/j.icarus.2017.12.046},
archivePrefix = {arXiv},
       eprint = {1801.03756},
 primaryClass = {astro-ph.EP},
       adsurl = {https://ui.adsabs.harvard.edu/abs/2018Icar..305..262M}
}

@ARTICLE{Li2020,
       author = {{Li}, Zhi and {Caracas}, Razvan and {Soubiran}, Fran{\c{c}}ois},
        title = "{Partial core vaporization during Giant Impacts inferred from the entropy and the critical point of iron}",
      journal = {Earth and Planetary Science Letters},
         year = 2020,
        month = oct,
       volume = {547},
          eid = {116463},
        pages = {116463},
          doi = {10.1016/j.epsl.2020.116463},
       adsurl = {https://ui.adsabs.harvard.edu/abs/2020E&PSL.54716463L}
}

@Article{Schlichting2012,
      author        = {{Schlichting}, Hilke E. and {Warren}, Paul H. and {Yin}, Qing-Zhu},
      journal       = {The Astrophysical Journal},
      title         = {{The Last Stages of Terrestrial Planet Formation: Dynamical Friction and the Late Veneer}},
      year          = {2012},
      month         = jun,
      number        = {1},
      pages         = {8},
      volume        = {752},
      adsurl        = {https://ui.adsabs.harvard.edu/abs/2012ApJ...752....8S},
      archiveprefix = {arXiv},
      doi           = {10.1088/0004-637X/752/1/8},
      eid           = {8},
      eprint        = {1202.6372},
      primaryclass  = {astro-ph.EP},
}

@ARTICLE{Bottke2010,
       author = {{Bottke}, William F. and {Walker}, Richard J. and {Day}, James M.~D. and {Nesvorny}, David and {Elkins-Tanton}, Linda},
        title = "{Stochastic Late Accretion to Earth, the Moon, and Mars}",
      journal = {Science},
         year = 2010,
        month = dec,
       volume = {330},
       number = {6010},
        pages = {1527},
          doi = {10.1126/science.1196874},
       adsurl = {https://ui.adsabs.harvard.edu/abs/2010Sci...330.1527B}
}

@ARTICLE{Raymond2013,
       author = {{Raymond}, Sean N. and {Schlichting}, Hilke E. and {Hersant}, Franck and {Selsis}, Franck},
        title = "{Dynamical and collisional constraints on a stochastic late veneer on the terrestrial planets}",
      journal = {Icarus},
         year = 2013,
        month = sep,
       volume = {226},
       number = {1},
        pages = {671-681},
          doi = {10.1016/j.icarus.2013.06.019},
archivePrefix = {arXiv},
       eprint = {1306.4325},
 primaryClass = {astro-ph.EP},
       adsurl = {https://ui.adsabs.harvard.edu/abs/2013Icar..226..671R}
}

@article{KorenagaMarchi2023,
author = {Jun Korenaga  and Simone Marchi },
title = {Vestiges of impact-driven three-phase mixing in the chemistry and structure of Earth’s mantle},
journal = {Proceedings of the National Academy of Sciences},
volume = {120},
number = {43},
pages = {e2309181120},
year = {2023},
doi = {10.1073/pnas.2309181120},
URL = {https://www.pnas.org/doi/abs/10.1073/pnas.2309181120},
eprint = {https://www.pnas.org/doi/pdf/10.1073/pnas.2309181120},
}

@ARTICLE{Genda2017,
       author = {{Genda}, H. and {Brasser}, R. and {Mojzsis}, S.~J.},
        title = "{The terrestrial late veneer from core disruption of a lunar-sized impactor}",
      journal = {Earth and Planetary Science Letters},
         year = 2017,
        month = dec,
       volume = {480},
        pages = {25-32},
          doi = {10.1016/j.epsl.2017.09.041},
archivePrefix = {arXiv},
       eprint = {1709.07554},
 primaryClass = {astro-ph.EP},
       adsurl = {https://ui.adsabs.harvard.edu/abs/2017E&PSL.480...25G}
}

@Article{Day2021,
  author   = {{Day}, James M.~D. and {Paquet}, Marine and {Timothy Jull}, A.~J.},
  journal  = {Meteoritics \& Planetary Science},
  title    = {{Temporally limited late accretion after core formation in the Moon}},
  year     = {2021},
  month    = apr,
  number   = {4},
  pages    = {683-699},
  volume   = {56},
  adsurl   = {https://ui.adsabs.harvard.edu/abs/2021M&PS...56..683D},
  doi      = {10.1111/maps.13646},
}

@ARTICLE{Marchi2018,
       author = {{Marchi}, S. and {Canup}, R.~M. and {Walker}, R.~J.},
        title = "{Heterogeneous delivery of silicate and metal to the Earth by large planetesimals}",
      journal = {Nature Geoscience},
         year = 2018,
        month = dec,
       volume = {11},
       number = {1},
        pages = {77-81},
          doi = {10.1038/s41561-017-0022-3},
       adsurl = {https://ui.adsabs.harvard.edu/abs/2018NatGe..11...77M}
}

@ARTICLE{Rubie2016,
       author = {{Rubie}, David C. and {Laurenz}, Vera and {Jacobson}, Seth A. and {Morbidelli}, Alessandro and {Palme}, Herbert and {Vogel}, Antje K. and {Frost}, Daniel J.},
        title = "{Highly siderophile elements were stripped from Earth{\textquoteright}s mantle by iron sulfide segregation}",
      journal = {Science},
         year = 2016,
        month = sep,
       volume = {353},
       number = {6304},
        pages = {1141-1144},
          doi = {10.1126/science.aaf6919},
archivePrefix = {arXiv},
       eprint = {1609.04751},
 primaryClass = {astro-ph.EP},
       adsurl = {https://ui.adsabs.harvard.edu/abs/2016Sci...353.1141R}
}

@ARTICLE{Day2007,
       author = {{Day}, James M.~D. and {Pearson}, D. Graham and {Taylor}, Lawrence A.},
        title = "{Highly Siderophile Element Constraints on Accretion and Differentiation of the Earth-Moon System}",
      journal = {Science},
         year = 2007,
        month = jan,
       volume = {315},
       number = {5809},
        pages = {217},
          doi = {10.1126/science.1133355},
       adsurl = {https://ui.adsabs.harvard.edu/abs/2007Sci...315..217D}
}

@ARTICLE{Walker2009,
       author = {{Walker}, Richard J.},
        title = "{Highly siderophile elements in the Earth, Moon and Mars: Update and implications for planetary accretion and differentiation}",
      journal = {Chemie der Erde / Geochemistry},
         year = 2009,
        month = jun,
       volume = {69},
       number = {2},
        pages = {101-125},
          doi = {10.1016/j.chemer.2008.10.001},
       adsurl = {https://ui.adsabs.harvard.edu/abs/2009ChEG...69..101W}
}

@ARTICLE{Mann2012,
       author = {{Mann}, Ute and {Frost}, Daniel J. and {Rubie}, David C. and {Becker}, Harry and {Aud{\'e}tat}, Andreas},
        title = "{Partitioning of Ru, Rh, Pd, Re, Ir and Pt between liquid metal and silicate at high pressures and high temperatures - Implications for the origin of highly siderophile element concentrations in the Earth's mantle}",
      journal = {Geochimica et Cosmochimica Acta},
         year = 2012,
        month = may,
       volume = {84},
        pages = {593-613},
          doi = {10.1016/j.gca.2012.01.026},
       adsurl = {https://ui.adsabs.harvard.edu/abs/2012GeCoA..84..593M}
}

@ARTICLE{DayWalker2015,
       author = {{Day}, James M.~D. and {Walker}, Richard J.},
        title = "{Highly siderophile element depletion in the Moon}",
      journal = {Earth and Planetary Science Letters},
         year = 2015,
        month = aug,
       volume = {423},
        pages = {114-124},
          doi = {10.1016/j.epsl.2015.05.001},
       adsurl = {https://ui.adsabs.harvard.edu/abs/2015E&PSL.423..114D}
}

@ARTICLE{Ryder2002,
       author = {{Ryder}, Graham},
        title = "{Mass flux in the ancient Earth-Moon system and benign implications for the origin of life on Earth}",
      journal = {Journal of Geophysical Research (Planets)},
         year = 2002,
        month = apr,
       volume = {107},
       number = {E4},
          eid = {5022},
        pages = {5022},
          doi = {10.1029/2001JE001583},
       adsurl = {https://ui.adsabs.harvard.edu/abs/2002JGRE..107.5022R}
}

@ARTICLE{ElkinsTanton2011,
       author = {{Elkins-Tanton}, Linda T. and {Burgess}, Seth and {Yin}, Qing-Zhu},
        title = "{The lunar magma ocean: Reconciling the solidification process with lunar petrology and geochronology}",
      journal = {Earth and Planetary Science Letters},
         year = 2011,
        month = apr,
       volume = {304},
       number = {3-4},
        pages = {326-336},
          doi = {10.1016/j.epsl.2011.02.004},
       adsurl = {https://ui.adsabs.harvard.edu/abs/2011E&PSL.304..326E}
}

@ARTICLE{Engels2024,
       author = {{Engels}, T. and {Monteux}, J. and {Boyet}, M. and {Bouhifd}, M.~A.},
        title = "{Large impacts and their contribution to the water budget of the Early Moon}",
      journal = {Icarus},
         year = 2024,
        month = aug,
       volume = {418},
          eid = {116124},
        pages = {116124},
          doi = {10.1016/j.icarus.2024.116124},
       adsurl = {https://ui.adsabs.harvard.edu/abs/2024Icar..41816124E}
}

@ARTICLE{Brasser2016,
       author = {{Brasser}, R. and {Mojzsis}, S.~J. and {Werner}, S.~C. and {Matsumura}, S. and {Ida}, S.},
        title = "{Late veneer and late accretion to the terrestrial planets}",
      journal = {Earth and Planetary Science Letters},
         year = 2016,
        month = dec,
       volume = {455},
        pages = {85-93},
          doi = {10.1016/j.epsl.2016.09.013},
archivePrefix = {arXiv},
       eprint = {1609.01785},
 primaryClass = {astro-ph.EP},
       adsurl = {https://ui.adsabs.harvard.edu/abs/2016E&PSL.455...85B}
}

@ARTICLE{Weidenschilling2011,
       author = {{Weidenschilling}, S.~J.},
        title = "{Initial sizes of planetesimals and accretion of the asteroids}",
      journal = {Icarus},
         year = 2011,
        month = aug,
       volume = {214},
       number = {2},
        pages = {671-684},
          doi = {10.1016/j.icarus.2011.05.024},
       adsurl = {https://ui.adsabs.harvard.edu/abs/2011Icar..214..671W}
}

@ARTICLE{Itcovitz2023,
       author = {{Itcovitz}, Jonathan P. and {Rae}, Auriol S.~P. and {Davison}, Thomas M. and {Collins}, Gareth S. and {Shorttle}, Oliver},
        title = "{The Distribution of Impactor Core Material During Large Impacts on Earth-like Planets}",
      journal = {The Planetary Science Journal},
         year = 2024,
        month = apr,
       volume = {5},
       number = {4},
          eid = {90},
        pages = {90},
          doi = {10.3847/PSJ/ad2ea4},
archivePrefix = {arXiv},
       eprint = {2312.12132},
 primaryClass = {astro-ph.EP},
       adsurl = {https://ui.adsabs.harvard.edu/abs/2024PSJ.....5...90I}
}

@ARTICLE{CitronStewart2022,
       author = {{Citron}, Robert I. and {Stewart}, Sarah T.},
        title = "{Large Impacts onto the Early Earth: Planetary Sterilization and Iron Delivery}",
      journal = {The Planetary Science Journal},
         year = 2022,
        month = may,
       volume = {3},
       number = {5},
          eid = {116},
        pages = {116},
          doi = {10.3847/PSJ/ac66e8},
archivePrefix = {arXiv},
       eprint = {2201.09349},
 primaryClass = {astro-ph.EP},
       adsurl = {https://ui.adsabs.harvard.edu/abs/2022PSJ.....3..116C}
}

@ARTICLE{ScottKrot2003,
       author = {{Scott}, E.~R.~D. and {Krot}, A.~N.},
        title = "{Chondrites and their Components}",
      journal = {Treatise on Geochemistry},
         year = 2003,
        month = dec,
       volume = {1},
        pages = {711},
          doi = {10.1016/B0-08-043751-6/01145-2},
       adsurl = {https://ui.adsabs.harvard.edu/abs/2003TrGeo...1..143S}
}

@ARTICLE{Landeau2021,
       author = {{Landeau}, Maylis and {Deguen}, Renaud and {Phillips}, Dominic and {Neufeld}, Jerome A. and {Lherm}, Victor and {Dalziel}, Stuart B.},
        title = "{Metal-silicate mixing by large Earth-forming impacts}",
      journal = {Earth and Planetary Science Letters},
         year = 2021,
        month = jun,
       volume = {564},
          eid = {116888},
        pages = {116888},
          doi = {10.1016/j.epsl.2021.116888},
       adsurl = {https://ui.adsabs.harvard.edu/abs/2021E&PSL.56416888L}
}

@ARTICLE{DahlStevenson2010,
       author = {{Dahl}, Tais W. and {Stevenson}, David J.},
        title = "{Turbulent mixing of metal and silicate during planet accretion {\textemdash} And interpretation of the Hf-W chronometer}",
      journal = {Earth and Planetary Science Letters},
         year = 2010,
        month = jun,
       volume = {295},
       number = {1-2},
        pages = {177-186},
          doi = {10.1016/j.epsl.2010.03.038},
       adsurl = {https://ui.adsabs.harvard.edu/abs/2010E&PSL.295..177D}
}

@ARTICLE{Deguen2014,
       author = {{Deguen}, Renaud and {Landeau}, Maylis and {Olson}, Peter},
        title = "{Turbulent metal-silicate mixing, fragmentation, and equilibration in magma oceans}",
      journal = {Earth and Planetary Science Letters},
         year = 2014,
        month = apr,
       volume = {391},
        pages = {274-287},
          doi = {10.1016/j.epsl.2014.02.007},
archivePrefix = {arXiv},
       eprint = {1402.1666},
 primaryClass = {physics.geo-ph},
       adsurl = {https://ui.adsabs.harvard.edu/abs/2014E&PSL.391..274D}
}

@ARTICLE{Jackson2023,
       author = {{Jackson}, Alan P. and {Perera}, Viranga and {Gabriel}, Travis S.~J.},
        title = "{Impact Generation of Holes in the Early Lunar Crust: Scaling Relations}",
      journal = {Journal of Geophysical Research (Planets)},
         year = 2023,
        month = apr,
       volume = {128},
       number = {4},
          eid = {e2022JE007498},
        pages = {e2022JE007498},
          doi = {10.1029/2022JE007498},
       adsurl = {https://ui.adsabs.harvard.edu/abs/2023JGRE..12807498J}
}

@ARTICLE{Canup2008,
       author = {{Canup}, Robin M.},
        title = "{Lunar-forming collisions with pre-impact rotation}",
      journal = {Icarus},
         year = 2008,
        month = aug,
       volume = {196},
       number = {2},
        pages = {518-538},
          doi = {10.1016/j.icarus.2008.03.011},
       adsurl = {https://ui.adsabs.harvard.edu/abs/2008Icar..196..518C}
}

@ARTICLE{Lock2018,
       author = {{Lock}, Simon J. and {Stewart}, Sarah T. and {Petaev}, Michail I. and {Leinhardt}, Zo{\"e} and {Mace}, Mia T. and {Jacobsen}, Stein B. and {Cuk}, Matija},
        title = "{The Origin of the Moon Within a Terrestrial Synestia}",
      journal = {Journal of Geophysical Research (Planets)},
         year = 2018,
        month = apr,
       volume = {123},
       number = {4},
        pages = {910-951},
          doi = {10.1002/2017JE005333},
archivePrefix = {arXiv},
       eprint = {1802.10223},
 primaryClass = {astro-ph.EP},
       adsurl = {https://ui.adsabs.harvard.edu/abs/2018JGRE..123..910L}
}

@ARTICLE{Brasser2020,
       author = {{Brasser}, R. and {Werner}, S.~C. and {Mojzsis}, S.~J.},
        title = "{Impact bombardment chronology of the terrestrial planets from 4.5 Ga to 3.5 Ga}",
      journal = {Icarus},
         year = 2020,
        month = mar,
       volume = {338},
          eid = {113514},
        pages = {113514},
          doi = {10.1016/j.icarus.2019.113514},
archivePrefix = {arXiv},
       eprint = {1910.11282},
 primaryClass = {astro-ph.EP},
       adsurl = {https://ui.adsabs.harvard.edu/abs/2020Icar..33813514B}
}

@ARTICLE{Nimmo2024,
       author = {{Nimmo}, Francis and {Kleine}, Thorsten and {Morbidelli}, Alessandro},
        title = "{Tidally driven remelting around 4.35 billion years ago indicates the Moon is old}",
      journal = {Nature},
         year = 2024,
        month = dec,
       volume = {636},
       number = {8043},
        pages = {598-602},
          doi = {10.1038/s41586-024-08231-0},
       adsurl = {https://ui.adsabs.harvard.edu/abs/2024Natur.636..598N}
}

@ARTICLE{NakajimaStevenson2015,
       author = {{Nakajima}, Miki and {Stevenson}, David J.},
        title = "{Melting and mixing states of the Earth's mantle after the Moon-forming impact}",
      journal = {Earth and Planetary Science Letters},
         year = 2015,
        month = oct,
       volume = {427},
        pages = {286-295},
          doi = {10.1016/j.epsl.2015.06.023},
archivePrefix = {arXiv},
       eprint = {1506.04853},
 primaryClass = {astro-ph.EP},
       adsurl = {https://ui.adsabs.harvard.edu/abs/2015E&PSL.427..286N}
}

@article{Maller2024,
    title = {Condition for metal fragmentation during Earth-forming collisions},
    journal = {Physics of the Earth and Planetary Interiors},
    volume = {352},
    pages = {107199},
    year = {2024},
    issn = {0031-9201},
    doi = {https://doi.org/10.1016/j.pepi.2024.107199},
    url = {https://www.sciencedirect.com/science/article/pii/S0031920124000578},
    author = {Augustin Maller and Maylis Landeau and Laetitia Allibert and Sébastien Charnoz},
}

@ARTICLE{SalvadorSamuel2023,
       author = {{Salvador}, Arnaud and {Samuel}, Henri},
        title = "{Convective outgassing efficiency in planetary magma oceans: Insights from computational fluid dynamics}",
      journal = {Icarus},
         year = 2023,
        month = jan,
       volume = {390},
          eid = {115265},
        pages = {115265},
          doi = {10.1016/j.icarus.2022.115265},
archivePrefix = {arXiv},
       eprint = {2209.06199},
 primaryClass = {astro-ph.EP},
       adsurl = {https://ui.adsabs.harvard.edu/abs/2023Icar..39015265S}
}

@ARTICLE{Day2016,
       author = {{Day}, James M.~D. and {Brandon}, Alan D. and {Walker}, Richard J.},
        title = "{Highly Siderophile Elements in Earth, Mars, the Moon, and Asteroids}",
      journal = {Reviews in Mineralogy and Geochemistry},
         year = 2016,
        month = jan,
       volume = {81},
       number = {1},
        pages = {161-238},
          doi = {10.2138/rmg.2016.81.04},
       adsurl = {https://ui.adsabs.harvard.edu/abs/2016RvMG...81..161D}
}

@ARTICLE{TonksMelosh1993,
       author = {{Tonks}, W. Brian and {Melosh}, H. Jay},
        title = "{Magma ocean formation due to giant impacts}",
      journal = {Journal of Geophysical Research},
         year = 1993,
        month = mar,
       volume = {98},
       number = {E3},
        pages = {5319-5333},
          doi = {10.1029/92JE02726},
       adsurl = {https://ui.adsabs.harvard.edu/abs/1993JGR....98.5319T}
}

@ARTICLE{Korenaga2018,
       author = {{Korenaga}, Jun},
        title = "{Crustal evolution and mantle dynamics through Earth history}",
      journal = {Philosophical Transactions of the Royal Society of London Series A},
         year = 2018,
        month = nov,
       volume = {376},
       number = {2132},
          eid = {20170408},
        pages = {20170408},
          doi = {10.1098/rsta.2017.0408},
       adsurl = {https://ui.adsabs.harvard.edu/abs/2018RSPTA.37670408K}
}

@ARTICLE{Nakajima2021,
       author = {{Nakajima}, Miki and {Golabek}, Gregor J. and {W{\"u}nnemann}, Kai and {Rubie}, David C. and {Burger}, Christoph and {Melosh}, Henry J. and {Jacobson}, Seth A. and {Manske}, Lukas and {Hull}, Scott D.},
        title = "{Scaling laws for the geometry of an impact-induced magma ocean}",
      journal = {Earth and Planetary Science Letters},
         year = 2021,
        month = aug,
       volume = {568},
          eid = {116983},
        pages = {116983},
          doi = {10.1016/j.epsl.2021.116983},
archivePrefix = {arXiv},
       eprint = {2004.04269},
 primaryClass = {astro-ph.EP},
       adsurl = {https://ui.adsabs.harvard.edu/abs/2021E&PSL.56816983N}
}

@ARTICLE{Wieczorek2013,
       author = {{Wieczorek}, Mark A. and {Neumann}, Gregory A. and {Nimmo}, Francis and {Kiefer}, Walter S. and {Taylor}, G. Jeffrey and {Melosh}, H. Jay and {Phillips}, Roger J. and {Solomon}, Sean C. and {Andrews-Hanna}, Jeffrey C. and {Asmar}, Sami W. and {Konopliv}, Alexander S. and {Lemoine}, Frank G. and {Smith}, David E. and {Watkins}, Michael M. and {Williams}, James G. and {Zuber}, Maria T.},
        title = "{The Crust of the Moon as Seen by GRAIL}",
      journal = {Science},
         year = 2013,
        month = feb,
       volume = {339},
       number = {6120},
        pages = {671-675},
          doi = {10.1126/science.1231530},
       adsurl = {https://ui.adsabs.harvard.edu/abs/2013Sci...339..671W}
}

@ARTICLE{Bottke2005,
       author = {{Bottke}, William F. and {Durda}, Daniel D. and {Nesvorn{\'y}}, David and {Jedicke}, Robert and {Morbidelli}, Alessandro and {Vokrouhlick{\'y}}, David and {Levison}, Hal},
        title = "{The fossilized size distribution of the main asteroid belt}",
      journal = {Icarus},
         year = 2005,
        month = may,
       volume = {175},
       number = {1},
        pages = {111-140},
          doi = {10.1016/j.icarus.2004.10.026},
       adsurl = {https://ui.adsabs.harvard.edu/abs/2005Icar..175..111B}
}

@ARTICLE{Salvador2017,
       author = {{Salvador}, A. and {Massol}, H. and {Davaille}, A. and {Marcq}, E. and {Sarda}, P. and {Chassefi{\`e}re}, E.},
        title = "{The relative influence of H$_{2}$O and CO$_{2}$ on the primitive surface conditions and evolution of rocky planets}",
      journal = {Journal of Geophysical Research (Planets)},
         year = 2017,
        month = jul,
       volume = {122},
       number = {7},
        pages = {1458-1486},
          doi = {10.1002/2017JE005286},
       adsurl = {https://ui.adsabs.harvard.edu/abs/2017JGRE..122.1458S}
}

@ARTICLE{KendallMelosh2016,
       author = {{Kendall}, Jordan D. and {Melosh}, H.~J.},
        title = "{Differentiated planetesimal impacts into a terrestrial magma ocean: Fate of the iron core}",
      journal = {Earth and Planetary Science Letters},
         year = 2016,
        month = aug,
       volume = {448},
        pages = {24-33},
          doi = {10.1016/j.epsl.2016.05.012},
       adsurl = {https://ui.adsabs.harvard.edu/abs/2016E&PSL.448...24K}
}

@ARTICLE{Lissauer1988,
       author = {{Lissauer}, Jack J. and {Squyres}, Steven W. and {Hartmann}, William K.},
        title = "{Bombardment history of the Saturn system.}",
      journal = {Journal of Geophysical Research},
         year = 1988,
        month = nov,
       volume = {93},
        pages = {13776-13804},
          doi = {10.1029/JB093iB11p13776},
       adsurl = {https://ui.adsabs.harvard.edu/abs/1988JGR....9313776L}
}

@ARTICLE{Samuel2012,
       author = {{Samuel}, Henri},
        title = "{A re-evaluation of metal diapir breakup and equilibration in terrestrial magma oceans}",
      journal = {Earth and Planetary Science Letters},
         year = 2012,
        month = jan,
       volume = {313},
        pages = {105-114},
          doi = {10.1016/j.epsl.2011.11.001},
       adsurl = {https://ui.adsabs.harvard.edu/abs/2012E&PSL.313..105S}
}

@ARTICLE{LavorelLeBars2009,
       author = {{Lavorel}, G. and {Le Bars}, M.},
        title = "{Sedimentation of particles in a vigorously convecting fluid}",
      journal = {Physical Review E},
         year = 2009,
        month = oct,
       volume = {80},
       number = {4},
          eid = {046324},
        pages = {046324},
          doi = {10.1103/PhysRevE.80.046324},
       adsurl = {https://ui.adsabs.harvard.edu/abs/2009PhRvE..80d6324L}
}

@ARTICLE{Freitas2021,
       author = {{Freitas}, D. and {Monteux}, J. and {Andrault}, D. and {Manthilake}, G. and {Mathieu}, A. and {Schiavi}, F. and {Cluzel}, N.},
        title = "{Thermal conductivities of solid and molten silicates: Implications for dynamos in mercury-like proto-planets}",
      journal = {Physics of the Earth and Planetary Interiors},
         year = 2021,
        month = mar,
       volume = {312},
          eid = {106655},
        pages = {106655},
          doi = {10.1016/j.pepi.2021.106655},
       adsurl = {https://ui.adsabs.harvard.edu/abs/2021PEPI..31206655F}
}

@ARTICLE{Durda1998,
       author = {{Durda}, Daniel D. and {Greenberg}, Richard and {Jedicke}, Robert},
        title = "{Collisional Models and Scaling Laws: A New Interpretation of the Shape of the Main-Belt Asteroid Size Distribution}",
      journal = {Icarus},
         year = 1998,
        month = oct,
       volume = {135},
       number = {2},
        pages = {431-440},
          doi = {10.1006/icar.1998.5960},
       adsurl = {https://ui.adsabs.harvard.edu/abs/1998Icar..135..431D}
}

@ARTICLE{Morbidelli2009,
       author = {{Morbidelli}, Alessandro and {Bottke}, William F. and {Nesvorn{\'y}}, David and {Levison}, Harold F.},
        title = "{Asteroids were born big}",
      journal = {Icarus},
         year = 2009,
        month = dec,
       volume = {204},
       number = {2},
        pages = {558-573},
          doi = {10.1016/j.icarus.2009.07.011},
archivePrefix = {arXiv},
       eprint = {0907.2512},
 primaryClass = {astro-ph.EP},
       adsurl = {https://ui.adsabs.harvard.edu/abs/2009Icar..204..558M}
}

@ARTICLE{Dohnanyi1969,
       author = {{Dohnanyi}, J.~S.},
        title = "{Collisional Model of Asteroids and Their Debris}",
      journal = {Journal of Geophysical Research},
         year = 1969,
        month = may,
       volume = {74},
        pages = {2531-2554},
          doi = {10.1029/JB074i010p02531},
       adsurl = {https://ui.adsabs.harvard.edu/abs/1969JGR....74.2531D}
}

@ARTICLE{Gladman2009,
       author = {{Gladman}, Brett J. and {Davis}, Donald R. and {Neese}, Carol and {Jedicke}, Robert and {Williams}, Gareth and {Kavelaars}, J.~J. and {Petit}, Jean-Marc and {Scholl}, Hans and {Holman}, Matthew and {Warrington}, Ben and {Esquerdo}, Gil and {Tricarico}, Pasquale},
        title = "{On the asteroid belt's orbital and size distribution}",
      journal = {Icarus},
         year = 2009,
        month = jul,
       volume = {202},
       number = {1},
        pages = {104-118},
          doi = {10.1016/j.icarus.2009.02.012},
       adsurl = {https://ui.adsabs.harvard.edu/abs/2009Icar..202..104G}
}

@ARTICLE{Masiero2011,
       author = {{Masiero}, Joseph R. and {Mainzer}, A.~K. and {Grav}, T. and {Bauer}, J.~M. and {Cutri}, R.~M. and {Dailey}, J. and {Eisenhardt}, P.~R.~M. and {McMillan}, R.~S. and {Spahr}, T.~B. and {Skrutskie}, M.~F. and {Tholen}, D. and {Walker}, R.~G. and {Wright}, E.~L. and {DeBaun}, E. and {Elsbury}, D. and {Gautier}, T., IV and {Gomillion}, S. and {Wilkins}, A.},
        title = "{Main Belt Asteroids with WISE/NEOWISE. I. Preliminary Albedos and Diameters}",
      journal = {The Astrophysical Journal},
         year = 2011,
        month = nov,
       volume = {741},
       number = {2},
          eid = {68},
        pages = {68},
          doi = {10.1088/0004-637X/741/2/68},
archivePrefix = {arXiv},
       eprint = {1109.4096},
 primaryClass = {astro-ph.EP},
       adsurl = {https://ui.adsabs.harvard.edu/abs/2011ApJ...741...68M}
}

@ARTICLE{Simon2016,
       author = {{Simon}, Jacob B. and {Armitage}, Philip J. and {Li}, Rixin and {Youdin}, Andrew N.},
        title = "{The Mass and Size Distribution of Planetesimals Formed by the Streaming Instability. I. The Role of Self-gravity}",
      journal = {The Astrophysical Journal},
         year = 2016,
        month = may,
       volume = {822},
       number = {1},
          eid = {55},
        pages = {55},
          doi = {10.3847/0004-637X/822/1/55},
archivePrefix = {arXiv},
       eprint = {1512.00009},
 primaryClass = {astro-ph.SR},
       adsurl = {https://ui.adsabs.harvard.edu/abs/2016ApJ...822...55S}
}

@ARTICLE{Johansen2015,
       author = {{Johansen}, Anders and {Mac Low}, Mordecai-Mark and {Lacerda}, Pedro and {Bizzarro}, Martin},
        title = "{Growth of asteroids, planetary embryos, and Kuiper belt objects by chondrule accretion}",
      journal = {Science Advances},
         year = 2015,
        month = apr,
       volume = {1},
          eid = {1500109},
        pages = {1500109},
          doi = {10.1126/sciadv.1500109},
archivePrefix = {arXiv},
       eprint = {1503.07347},
 primaryClass = {astro-ph.EP},
       adsurl = {https://ui.adsabs.harvard.edu/abs/2015SciA....1E0109J}
}

@ARTICLE{Touboul2015,
       author = {{Touboul}, Mathieu and {Puchtel}, Igor S. and {Walker}, Richard J.},
        title = "{Tungsten isotopic evidence for disproportional late accretion to the Earth and Moon}",
      journal = {Nature},
         year = 2015,
        month = apr,
       volume = {520},
       number = {7548},
        pages = {530-533},
          doi = {10.1038/nature14355},
       adsurl = {https://ui.adsabs.harvard.edu/abs/2015Natur.520..530T}
}

@ARTICLE{Kruijer2015,
       author = {{Kruijer}, Thomas S. and {Kleine}, Thorsten and {Fischer-G{\"o}dde}, Mario and {Sprung}, Peter},
        title = "{Lunar tungsten isotopic evidence for the late veneer}",
      journal = {Nature},
         year = 2015,
        month = apr,
       volume = {520},
       number = {7548},
        pages = {534-537},
          doi = {10.1038/nature14360},
       adsurl = {https://ui.adsabs.harvard.edu/abs/2015Natur.520..534K}
}

@ARTICLE{LhermDeguen2018,
       author = {{Lherm}, V. and {Deguen}, R.},
        title = "{Small-Scale Metal/Silicate Equilibration During Core Formation: The Influence of Stretching Enhanced Diffusion on Mixing}",
      journal = {Journal of Geophysical Research (Solid Earth)},
         year = 2018,
        month = dec,
       volume = {123},
       number = {12},
        pages = {10,496-10,516},
          doi = {10.1029/2018JB016537},
archivePrefix = {arXiv},
       eprint = {1812.07855},
 primaryClass = {physics.geo-ph},
       adsurl = {https://ui.adsabs.harvard.edu/abs/2018JGRB..12310496L}
}

@ARTICLE{Ulvrova2011,
       author = {{Ulvrov{\'a}}, M. and {Coltice}, N. and {Ricard}, Y. and {Labrosse}, S. and {Dubuffet}, F. and {Vel{\'\i}Msk{\'y}}, J. and {{\r{A}} R{\'a}Mek}, O.},
        title = "{Compositional and thermal equilibration of particles, drops, and diapirs in geophysical flows}",
      journal = {Geochemistry, Geophysics, Geosystems},
         year = 2011,
        month = oct,
       volume = {12},
       number = {10},
          eid = {Q10014},
        pages = {Q10014},
          doi = {10.1029/2011GC003757},
       adsurl = {https://ui.adsabs.harvard.edu/abs/2011GGG....1210014U}
}

@ARTICLE{Albarede2013,
       author = {{Albar{\`e}de}, Francis and {Ballhaus}, Chris and {Blichert-Toft}, Janne and {Lee}, Cin-Ty and {Marty}, Bernard and {Moynier}, Fr{\'e}d{\'e}ric and {Yin}, Qing-Zhu},
        title = "{Asteroidal impacts and the origin of terrestrial and lunar volatiles}",
      journal = {Icarus},
         year = 2013,
        month = jan,
       volume = {222},
       number = {1},
        pages = {44-52},
          doi = {10.1016/j.icarus.2012.10.026},
       adsurl = {https://ui.adsabs.harvard.edu/abs/2013Icar..222...44A}
}

@ARTICLE{PierazzoMelosh2000,
       author = {{Pierazzo}, E. and {Melosh}, H.~J.},
        title = "{Hydrocode modeling of oblique impacts: The fate of the projectile}",
      journal = {Meteoritics \& Planetary Science},
         year = 2000,
        month = jan,
       volume = {35},
       number = {1},
        pages = {117-130},
          doi = {10.1111/j.1945-5100.2000.tb01979.x},
       adsurl = {https://ui.adsabs.harvard.edu/abs/2000M&PS...35..117P}
}

@ARTICLE{CatlingZahnle2020,
       author = {{Catling}, David C. and {Zahnle}, Kevin J.},
        title = "{The Archean atmosphere}",
      journal = {Science Advances},
         year = 2020,
        month = feb,
       volume = {6},
       number = {9},
          eid = {eaax1420},
        pages = {eaax1420},
          doi = {10.1126/sciadv.aax1420},
       adsurl = {https://ui.adsabs.harvard.edu/abs/2020SciA....6.1420C}
}

@ARTICLE{KaratoRamaMurthy1997,
       author = {{Karato}, Shun-ichiro and {Rama Murthy}, V.},
        title = "{Core formation and chemical equilibrium in the Earth{\textemdash}I. Physical considerations}",
      journal = {Physics of the Earth and Planetary Interiors},
         year = 1997,
        month = mar,
       volume = {100},
       number = {1},
        pages = {61-79},
          doi = {10.1016/S0031-9201(96)03232-3},
       adsurl = {https://ui.adsabs.harvard.edu/abs/1997PEPI..100...61K}
}

@ARTICLE{PotterCollins2013,
       author = {{Potter}, Ross W.~K. and {Collins}, Gareth S.},
        title = "{Numerical modeling of asteroid survivability and possible scenarios for the Morokweng crater-forming impact}",
      journal = {Meteoritics \& Planetary Science},
         year = 2013,
        month = may,
       volume = {48},
       number = {5},
        pages = {744-757},
          doi = {10.1111/maps.12098},
       adsurl = {https://ui.adsabs.harvard.edu/abs/2013M&PS...48..744P}
}

@article{Wunnemann2008,
    title = {Numerical modelling of impact melt production in porous rocks},
    journal = {Earth and Planetary Science Letters},
    volume = {269},
    number = {3},
    pages = {530-539},
    year = {2008},
    issn = {0012-821X},
    doi = {https://doi.org/10.1016/j.epsl.2008.03.007},
    url = {https://www.sciencedirect.com/science/article/pii/S0012821X08001660},
    author = {K. Wünnemann and G.S. Collins and G.R. Osinski},
}

@BOOK{Melosh1989,
       author = {{Melosh}, H.~J.},
        title = "{Impact cratering : a geologic process}",
         year = 1989,
       adsurl = {https://ui.adsabs.harvard.edu/abs/1989icgp.book.....M}
}

@ARTICLE{Chou1978,
       author = {{Chou}, C. -L.},
        title = "{Fractionation of Siderophile Elements in the Earth's Upper Mantle}",
      journal = {Lunar and Planetary Science Conference Proceedings},
         year = 1978,
        month = jan,
       volume = {1},
        pages = {219-230},
       adsurl = {https://ui.adsabs.harvard.edu/abs/1978LPSC....9..219C}
}

@ARTICLE{Allibert2023,
       author = {{Allibert}, L. and {Landeau}, M. and {R{\"o}hlen}, R. and {Maller}, A. and {Nakajima}, M. and {W{\"u}nnemann}, K.},
        title = "{Planetary Impacts: Scaling of Crater Depth From Subsonic to Supersonic Conditions}",
      journal = {Journal of Geophysical Research (Planets)},
         year = 2023,
        month = aug,
       volume = {128},
       number = {8},
          eid = {e2023JE007823},
        pages = {e2023JE007823},
          doi = {10.1029/2023JE007823},
       adsurl = {https://ui.adsabs.harvard.edu/abs/2023JGRE..12807823A}
}

@ARTICLE{Suer2021,
       author = {{Suer}, Terry-Ann and {Siebert}, Julien and {Remusat}, Laurent and {Day}, James M.~D. and {Borensztajn}, Stephan and {Doisneau}, Beatrice and {Fiquet}, Guillaume},
        title = "{Reconciling metal-silicate partitioning and late accretion in the Earth}",
      journal = {Nature Communications},
         year = 2021,
        month = jan,
       volume = {12},
          eid = {2913},
        pages = {2913},
          doi = {10.1038/s41467-021-23137-5},
       adsurl = {https://ui.adsabs.harvard.edu/abs/2021NatCo..12.2913S}
}

@ARTICLE{Righter2008,
       author = {{Righter}, K. and {Humayun}, M. and {Danielson}, L.},
        title = "{Partitioning of palladium at high pressures and temperatures during core formation}",
      journal = {Nature Geoscience},
         year = 2008,
        month = may,
       volume = {1},
       number = {5},
        pages = {321-323},
          doi = {10.1038/ngeo180},
       adsurl = {https://ui.adsabs.harvard.edu/abs/2008NatGe...1..321R}
}

@Article{Rimmer2019,
    author={Rimmer, P. B.
    and Shorttle, O.
    and Rugheimer, S.},
    title={Oxidised micrometeorites as evidence for low atmospheric pressure on the early Earth},
    journal={Geochemical Perspectives Letters},
    year={2019},
    volume={9},
    pages={38-42},
    url={https://www.geochemicalperspectivesletters.org/article1903}
}

@ARTICLE{Dale2012,
       author = {{Dale}, Christopher W. and {Burton}, Kevin W. and {Greenwood}, Richard C. and {Gannoun}, Abdelmouhcine and {Wade}, Jonathan and {Wood}, Bernard J. and {Pearson}, D. Graham},
        title = "{Late Accretion on the Earliest Planetesimals Revealed by the Highly Siderophile Elements}",
      journal = {Science},
         year = 2012,
        month = apr,
       volume = {336},
       number = {6077},
        pages = {72},
          doi = {10.1126/science.1214967},
       adsurl = {https://ui.adsabs.harvard.edu/abs/2012Sci...336...72D}
}

@ARTICLE{Wood1970,
       author = {{Wood}, J.~A. and {Dickey}, Jr., J.~S. and {Marvin}, Ursula B. and {Powell}, B.~N.},
        title = "{Lunar anorthosites and a geophysical model of the moon}",
      journal = {Geochimica et Cosmochimica Acta Supplement},
         year = 1970,
        month = jan,
       volume = {1},
        pages = {965},
       adsurl = {https://ui.adsabs.harvard.edu/abs/1970GeCAS...1..965W}
}

@ARTICLE{Elkins-Tanton2008,
       author = {{Elkins-Tanton}, L.~T.},
        title = "{Linked magma ocean solidification and atmospheric growth for Earth and Mars}",
      journal = {Earth and Planetary Science Letters},
         year = 2008,
        month = jul,
       volume = {271},
       number = {1-4},
        pages = {181-191},
          doi = {10.1016/j.epsl.2008.03.062},
       adsurl = {https://ui.adsabs.harvard.edu/abs/2008E&PSL.271..181E}
}

@ARTICLE{Kraus2015,
       author = {{Kraus}, Richard G. and {Root}, Seth and {Lemke}, Raymond W. and {Stewart}, Sarah T. and {Jacobsen}, Stein B. and {Mattsson}, Thomas R.},
        title = "{Impact vaporization of planetesimal cores in the late stages of planet formation}",
      journal = {Nature Geoscience},
         year = 2015,
        month = apr,
       volume = {8},
       number = {4},
        pages = {269-272},
          doi = {10.1038/ngeo2369},
       adsurl = {https://ui.adsabs.harvard.edu/abs/2015NatGe...8..269K}
}

@ARTICLE{Kleine2009,
       author = {{Kleine}, Thorsten and {Touboul}, Mathieu and {Bourdon}, Bernard and {Nimmo}, Francis and {Mezger}, Klaus and {Palme}, Herbert and {Jacobsen}, Stein B. and {Yin}, Qing-Zhu and {Halliday}, Alexander N.},
        title = "{Hf-W chronology of the accretion and early evolution of asteroids and terrestrial planets}",
      journal = {Geochimica et Cosmochimica Acta},
         year = 2009,
        month = sep,
       volume = {73},
       number = {17},
        pages = {5150-5188},
          doi = {10.1016/j.gca.2008.11.047},
       adsurl = {https://ui.adsabs.harvard.edu/abs/2009GeCoA..73.5150K}
}

@article{StixrudeKarki2010,
  title={Viscosity of MgSiO3 liquid at Earth’s mantle conditions: Implications for an early magma ocean},
  author={Karki, Bijaya B and Stixrude, Lars P},
  journal={Science},
  volume={328},
  number={5979},
  pages={740--742},
  year={2010},
  publisher={American Association for the Advancement of Science}
}

@article{Hinze1955,
  title={Fundamentals of the hydrodynamic mechanism of splitting in dispersion processes},
  author={Hinze, Julius O},
  journal={AIChE journal},
  volume={1},
  number={3},
  pages={289--295},
  year={1955},
  publisher={Wiley Online Library}
}

@article{Costa2005,
  title={Viscosity of high crystal content melts: dependence on solid fraction},
  author={Costa, Antonio},
  journal={Geophysical Research Letters},
  volume={32},
  number={22},
  year={2005},
  publisher={Wiley Online Library}
}

@article{Lejeune1995,
  title={Rheology of crystal-bearing silicate melts: An experimental study at high viscosities},
  author={Lejeune, Anne-Marie and Richet, Pascal},
  journal={Journal of Geophysical Research: Solid Earth},
  volume={100},
  number={B3},
  pages={4215--4229},
  year={1995},
  publisher={Wiley Online Library}
}

@article{Olson2008,
  title={Experiments on metal--silicate plumes and core formation},
  author={Olson, Peter and Weeraratne, Dayanthie},
  journal={Philosophical Transactions of the Royal Society A: Mathematical, Physical and Engineering Sciences},
  volume={366},
  number={1883},
  pages={4253--4271},
  year={2008},
  publisher={The Royal Society London}
}

@article{Fleck2018,
  title={Iron diapirs entrain silicates to the core and initiate thermochemical plumes},
  author={Fleck, JR and Rains, CL and Weeraratne, DS and Nguyen, CT and Brand, DM and Klein, SM and McGehee, JM and Rincon, JM and Martinez, Cintia and Olson, PL},
  journal={Nature communications},
  volume={9},
  number={1},
  pages={71},
  year={2018},
  publisher={Nature Publishing Group UK London}
}

@ARTICLE{Brandon2012,
       author = {{Brandon}, Alan D. and {Puchtel}, Igor S. and {Walker}, Richard J. and {Day}, James M.~D. and {Irving}, Anthony J. and {Taylor}, Lawrence A.},
        title = "{Evolution of the martian mantle inferred from the $^{187}$Re- $^{187}$Os isotope and highly siderophile element abundance systematics of shergottite meteorites}",
      journal = {Geochimica et Cosmochimica Acta},
         year = 2012,
        month = jan,
       volume = {76},
        pages = {206-235},
          doi = {10.1016/j.gca.2011.09.047},
       adsurl = {https://ui.adsabs.harvard.edu/abs/2012GeCoA..76..206B}
}

@ARTICLE{Brandon2000,
       author = {{Brandon}, A.~D. and {Walker}, R.~J. and {Morgan}, J.~W. and {Goles}, G.~G.},
        title = "{Re-Os isotopic evidence for early differentiation of the Martian mantle}",
      journal = {Geochimica et Cosmochimica Acta},
         year = 2000,
        month = dec,
       volume = {64},
       number = {23},
        pages = {4083-4095},
          doi = {10.1016/S0016-7037(00)00482-8},
       adsurl = {https://ui.adsabs.harvard.edu/abs/2000GeCoA..64.4083B}
}

@ARTICLE{Hamano2013,
       author = {{Hamano}, Keiko and {Abe}, Yutaka and {Genda}, Hidenori},
        title = "{Emergence of two types of terrestrial planet on solidification of magma ocean}",
      journal = {Nature},
         year = 2013,
        month = may,
       volume = {497},
       number = {7451},
        pages = {607-610},
          doi = {10.1038/nature12163},
       adsurl = {https://ui.adsabs.harvard.edu/abs/2013Natur.497..607H}
}

@ARTICLE{Boyet2005,
       author = {{Boyet}, M. and {Carlson}, R.~W.},
        title = "{$^{142}$Nd Evidence for Early (>4.53 Ga) Global Differentiation of the Silicate Earth}",
      journal = {Science},
         year = 2005,
        month = jul,
       volume = {309},
       number = {5734},
        pages = {576-581},
          doi = {10.1126/science.1113634},
       adsurl = {https://ui.adsabs.harvard.edu/abs/2005Sci...309..576B}
}

@ARTICLE{Kleine2005,
       author = {{Kleine}, Thorsten and {Palme}, Herbert and {Mezger}, Klaus and {Halliday}, Alex N.},
        title = "{Hf-W Chronometry of Lunar Metals and the Age and Early Differentiation of the Moon}",
      journal = {Science},
         year = 2005,
        month = dec,
       volume = {310},
       number = {5754},
        pages = {1671-1674},
          doi = {10.1126/science.1118842},
       adsurl = {https://ui.adsabs.harvard.edu/abs/2005Sci...310.1671K}
}

@ARTICLE{PahlevanMorbidelli2015,
       author = {{Pahlevan}, Kaveh and {Morbidelli}, Alessandro},
        title = "{Collisionless encounters and the origin of the lunar inclination}",
      journal = {Nature},
         year = 2015,
        month = nov,
       volume = {527},
       number = {7579},
        pages = {492-494},
          doi = {10.1038/nature16137},
archivePrefix = {arXiv},
       eprint = {1603.06515},
 primaryClass = {astro-ph.EP},
       adsurl = {https://ui.adsabs.harvard.edu/abs/2015Natur.527..492P}
}

@ARTICLE{Valley2014,
       author = {{Valley}, John W. and {Cavosie}, Aaron J. and {Ushikubo}, Takayuki and {Reinhard}, David A. and {Lawrence}, Daniel F. and {Larson}, David J. and {Clifton}, Peter H. and {Kelly}, Thomas F. and {Wilde}, Simon A. and {Moser}, Desmond E. and {Spicuzza}, Michael J.},
        title = "{Hadean age for a post-magma-ocean zircon confirmed by atom-probe tomography}",
      journal = {Nature Geoscience},
         year = 2014,
        month = mar,
       volume = {7},
       number = {3},
        pages = {219-223},
          doi = {10.1038/ngeo2075},
       adsurl = {https://ui.adsabs.harvard.edu/abs/2014NatGe...7..219V}
}

@ARTICLE{Marchi2014,
       author = {{Marchi}, S. and {Bottke}, W.~F. and {Elkins-Tanton}, L.~T. and {Bierhaus}, M. and {Wuennemann}, K. and {Morbidelli}, A. and {Kring}, D.~A.},
        title = "{Widespread mixing and burial of Earth's Hadean crust by asteroid impacts}",
      journal = {Nature},
         year = 2014,
        month = jul,
       volume = {511},
       number = {7511},
        pages = {578-582},
          doi = {10.1038/nature13539},
       adsurl = {https://ui.adsabs.harvard.edu/abs/2014Natur.511..578M}
}

@ARTICLE{Jacobson2014,
       author = {{Jacobson}, Seth A. and {Morbidelli}, Alessandro and {Raymond}, Sean N. and {O'Brien}, David P. and {Walsh}, Kevin J. and {Rubie}, David C.},
        title = "{Highly siderophile elements in Earth's mantle as a clock for the Moon-forming impact}",
      journal = {Nature},
         year = 2014,
        month = apr,
       volume = {508},
       number = {7494},
        pages = {84-87},
          doi = {10.1038/nature13172},
archivePrefix = {arXiv},
       eprint = {1504.01421},
 primaryClass = {astro-ph.EP},
       adsurl = {https://ui.adsabs.harvard.edu/abs/2014Natur.508...84J}
}

@ARTICLE{Woo2024,
       author = {{Woo}, J.~M.~Y. and {Nesvorn{\'y}}, D. and {Scora}, J. and {Morbidelli}, A.},
        title = "{Terrestrial planet formation from a ring: Long-term simulations accounting for the giant planet instability}",
      journal = {Icarus},
         year = 2024,
        month = jul,
       volume = {417},
          eid = {116109},
        pages = {116109},
          doi = {10.1016/j.icarus.2024.116109},
archivePrefix = {arXiv},
       eprint = {2404.17259},
 primaryClass = {astro-ph.EP},
       adsurl = {https://ui.adsabs.harvard.edu/abs/2024Icar..41716109W}
}

@ARTICLE{Touboul2007,
       author = {{Touboul}, M. and {Kleine}, T. and {Bourdon}, B. and {Palme}, H. and {Wieler}, R.},
        title = "{Late formation and prolonged differentiation of the Moon inferred from W isotopes in lunar metals}",
      journal = {Nature},
         year = 2007,
        month = dec,
       volume = {450},
       number = {7173},
        pages = {1206-1209},
          doi = {10.1038/nature06428},
       adsurl = {https://ui.adsabs.harvard.edu/abs/2007Natur.450.1206T}
}

@ARTICLE{Mojzsis2019,
       author = {{Mojzsis}, Stephen J. and {Brasser}, Ramon and {Kelly}, Nigel M. and {Abramov}, Oleg and {Werner}, Stephanie C.},
        title = "{Onset of Giant Planet Migration before 4480 Million Years Ago}",
      journal = {The Astrophysical Journal},
         year = 2019,
        month = aug,
       volume = {881},
       number = {1},
          eid = {44},
        pages = {44},
          doi = {10.3847/1538-4357/ab2c03},
archivePrefix = {arXiv},
       eprint = {1903.08825},
 primaryClass = {astro-ph.EP},
       adsurl = {https://ui.adsabs.harvard.edu/abs/2019ApJ...881...44M}
}

@article{Chopelas1992,
  title={Thermal expansivity in the lower mantle},
  author={Chopelas, A and Boehler, R},
  journal={Geophysical Research Letters},
  volume={19},
  number={19},
  pages={1983--1986},
  year={1992},
  publisher={Wiley Online Library}
}

@article{Miller1991,
	Address = {{2000 FLORIDA AVE NW, WASHINGTON, DC 20009}},
	Affiliation = {{CALTECH,DIV GEOL \& PLANETARY SCI,PASADENA,CA 91125. CALTECH,HELEN E LINDHURST LAB EXPTL GEOPHYS,PASADENA,CA 91125.}},
	Author = {Miller, G. H. and Stolper, E. M. and Ahrens, T. J.},
	Doc-Delivery-Number = {{FW819}},
	Doi = {{10.1029/91JB01204}},
	Issn = {{0148-0227}},
	Journal = {J. Geophys. Res.},
	Keywords-Plus = {{SINGLE-IMPACT HYPOTHESIS; SOLID-SOLUTION SERIES; LOWER-MANTLE; HIGH-PRESSURES; SILICATE LIQUIDS; DRY PERIDOTITE; MELT DENSITY; MAGMA OCEAN; 14 GPA; ELASTICITY}},
	Language = {{English}},
	Month = {{JUL 10}},
	Number = {{B7}},
	Number-Of-Cited-References = {{89}},
	Pages = {{11831-11848}},
	Publisher = {{AMER GEOPHYSICAL UNION}},
	Research-Areas = {{Geology}},
	Times-Cited = {{84}},
	Title = {The equation of state of a molten komatiite.1. Shock-wave compression to 36 GPa},
	Type = {{Article}},
	Unique-Id = {{ISI:A1991FW81900013}},
	Volume = {{96}},
	Web-Of-Science-Categories = {{Geosciences, Multidisciplinary}},
	Year = {{1991}}}

@article{Morard2013,
	Author = {Morard, G. and Siebert, J. and Andrault, D. and Guignot, N. and Garbarino, G. and Guyot, F. and Antonangeli, D.},
	Journal = {Earth Planet. Sci. Lett.},
	Pages = {169-178},
	Publisher = {Elsevier},
	Title = {The Earth's core composition from high pressure density measurements of liquid iron alloys},
	Volume = {373},
	Year = {2013}}

@article{Stebbins1984,
  title={Heat capacities and entropies of silicate liquids and glasses},
  author={Stebbins, JF and Carmichael, ISE and Moret, LK},
  journal={Contributions to mineralogy and petrology},
  volume={86},
  pages={131--148},
  year={1984},
  publisher={Springer}
}

@ARTICLE{Maurice2017,
       author = {{Maurice}, Maxime and {Tosi}, Nicola and {Samuel}, Henri and {Plesa}, Ana-Catalina and {H{\"u}ttig}, Christian and {Breuer}, Doris},
        title = "{Onset of solid-state mantle convection and mixing during magma ocean solidification}",
      journal = {Journal of Geophysical Research (Planets)},
         year = 2017,
        month = mar,
       volume = {122},
       number = {3},
        pages = {577-598},
          doi = {10.1002/2016JE005250},
       adsurl = {https://ui.adsabs.harvard.edu/abs/2017JGRE..122..577M}
}

@ARTICLE{deKoker2010,
       author = {{de Koker}, Nico},
        title = "{Thermal conductivity of MgO periclase at high pressure: Implications for the D\" region}",
      journal = {Earth and Planetary Science Letters},
         year = 2010,
        month = apr,
       volume = {292},
       number = {3-4},
        pages = {392-398},
          doi = {10.1016/j.epsl.2010.02.011},
       adsurl = {https://ui.adsabs.harvard.edu/abs/2010E&PSL.292..392D}
}

@article{Lebrun2013,
  title={Thermal evolution of an early magma ocean in interaction with the atmosphere},
  author={Lebrun, Thomas and Massol, H{\'e}l{\`e}ne and Chassefi{\`e}re, Eric and Davaille, Anne and Marcq, Emmanuel and Sarda, Philippe and Leblanc, Fran{\c{c}}ois and Brandeis, Genevi{\`e}ve},
  journal={Journal of Geophysical Research: Planets},
  volume={118},
  number={6},
  pages={1155--1176},
  year={2013},
  publisher={Wiley Online Library}
}

@ARTICLE{FischerGodde2020,
       author = {{Fischer-G{\"o}dde}, Mario and {Elfers}, Bo-Magnus and {M{\"u}nker}, Carsten and {Szilas}, Kristoffer and {Maier}, Wolfgang D. and {Messling}, Nils and {Morishita}, Tomoaki and {Van Kranendonk}, Martin and {Smithies}, Hugh},
        title = "{Ruthenium isotope vestige of Earth's pre-late-veneer mantle preserved in Archaean rocks}",
      journal = {Nature},
         year = 2020,
        month = mar,
       volume = {579},
       number = {7798},
        pages = {240-244},
          doi = {10.1038/s41586-020-2069-3},
       adsurl = {https://ui.adsabs.harvard.edu/abs/2020Natur.579..240F}
}

@ARTICLE{Marty2012,
       author = {{Marty}, Bernard},
        title = "{The origins and concentrations of water, carbon, nitrogen and noble gases on Earth}",
      journal = {Earth and Planetary Science Letters},
         year = 2012,
        month = jan,
       volume = {313},
        pages = {56-66},
          doi = {10.1016/j.epsl.2011.10.040},
archivePrefix = {arXiv},
       eprint = {1405.6336},
 primaryClass = {astro-ph.EP},
       adsurl = {https://ui.adsabs.harvard.edu/abs/2012E&PSL.313...56M}
}

@ARTICLE{Thomas2005,
       author = {{Thomas}, P.~C. and {Parker}, J. Wm. and {McFadden}, L.~A. and {Russell}, C.~T. and {Stern}, S.~A. and {Sykes}, M.~V. and {Young}, E.~F.},
        title = "{Differentiation of the asteroid Ceres as revealed by its shape}",
      journal = {Nature},
         year = 2005,
        month = sep,
       volume = {437},
       number = {7056},
        pages = {224-226},
          doi = {10.1038/nature03938},
       adsurl = {https://ui.adsabs.harvard.edu/abs/2005Natur.437..224T}
}

@ARTICLE{Schubert1996,
       author = {{Schubert}, Gerald and {Zhang}, Keke and {Kivelson}, Margaret G. and {Anderson}, John D.},
        title = "{The magnetic field and internal structure of Ganymede}",
      journal = {Nature},
         year = 1996,
        month = dec,
       volume = {384},
       number = {6609},
        pages = {544-545},
          doi = {10.1038/384544a0},
       adsurl = {https://ui.adsabs.harvard.edu/abs/1996Natur.384..544S}
}

@ARTICLE{Grewal2024,
       author = {{Grewal}, Damanveer S. and {Nie}, Nicole X. and {Zhang}, Bidong and {Izidoro}, Andre and {Asimow}, Paul D.},
        title = "{Accretion of the earliest inner Solar System planetesimals beyond the water snowline}",
      journal = {Nature Astronomy},
         year = 2024,
        month = mar,
       volume = {8},
        pages = {290-297},
          doi = {10.1038/s41550-023-02172-w},
archivePrefix = {arXiv},
       eprint = {2408.17032},
 primaryClass = {astro-ph.EP},
       adsurl = {https://ui.adsabs.harvard.edu/abs/2024NatAs...8..290G}
}

@ARTICLE{Rubie2003,
       author = {{Rubie}, D.~C. and {Melosh}, H.~J. and {Reid}, J.~E. and {Liebske}, C. and {Righter}, K.},
        title = "{Mechanisms of metal-silicate equilibration in the terrestrial magma ocean}",
      journal = {Earth and Planetary Science Letters},
         year = 2003,
        month = jan,
       volume = {205},
       number = {3-4},
        pages = {239-255},
          doi = {10.1016/S0012-821X(02)01044-0},
       adsurl = {https://ui.adsabs.harvard.edu/abs/2003E&PSL.205..239R}
}

@article{Clesi2020,
    title = {Dynamics of core-mantle separation: Influence of viscosity contrast and metal/silicate partition coefficients on the chemical equilibrium},
    journal = {Physics of the Earth and Planetary Interiors},
    volume = {306},
    pages = {106547},
    year = {2020},
    issn = {0031-9201},
    doi = {https://doi.org/10.1016/j.pepi.2020.106547},
    url = {https://www.sciencedirect.com/science/article/pii/S0031920119302456},
    author = {V. Clesi and J. Monteux and B. Qaddah and M. {Le Bars} and J.-B. Wacheul and M.A. Bouhifd},
}

@ARTICLE{WacheulLeBars2018,
       author = {{Wacheul}, Jean-Baptiste and {Le Bars}, Michael},
        title = "{Experiments on fragmentation and thermo-chemical exchanges during planetary core formation}",
      journal = {Physics of the Earth and Planetary Interiors},
         year = 2018,
        month = mar,
       volume = {276},
        pages = {134-144},
          doi = {10.1016/j.pepi.2017.05.018},
       adsurl = {https://ui.adsabs.harvard.edu/abs/2018PEPI..276..134W}
}

@ARTICLE{Andrault2011,
       author = {{Andrault}, Denis and {Bolfan-Casanova}, Nathalie and {Nigro}, Giacomo Lo and {Bouhifd}, Mohamed A. and {Garbarino}, Gaston and {Mezouar}, Mohamed},
        title = "{Solidus and liquidus profiles of chondritic mantle: Implication for melting of the Earth across its history}",
      journal = {Earth and Planetary Science Letters},
         year = 2011,
        month = apr,
       volume = {304},
       number = {1-2},
        pages = {251-259},
          doi = {10.1016/j.epsl.2011.02.006},
       adsurl = {https://ui.adsabs.harvard.edu/abs/2011E&PSL.304..251A}
}

@article{Ballmer2017,
    author = {Ballmer, Maxim D. and Lourenço, Diogo L. and Hirose, Kei and Caracas, Razvan and Nomura, Ryuichi},
    title = {Reconciling magma-ocean crystallization models with the present-day structure of the Earth's mantle},
    journal = {Geochemistry, Geophysics, Geosystems},
    volume = {18},
    number = {7},
    pages = {2785-2806},
    doi = {https://doi.org/10.1002/2017GC006917},
    url = {https://agupubs.onlinelibrary.wiley.com/doi/abs/10.1002/2017GC006917},
    eprint = {https://agupubs.onlinelibrary.wiley.com/doi/pdf/10.1002/2017GC006917},
    year = {2017}
}

@ARTICLE{Boukare2025,
       author = {{Boukar{\'e}}, Charles-{\'E}douard and {Badro}, James and {Samuel}, Henri},
        title = "{Solidification of Earth's mantle led inevitably to a basal magma ocean}",
      journal = {Nature},
         year = 2025,
        month = apr,
       volume = {640},
       number = {8057},
        pages = {114-119},
          doi = {10.1038/s41586-025-08701-z},
       adsurl = {https://ui.adsabs.harvard.edu/abs/2025Natur.640..114B}
}

@article{Labrosse2024,
     author = {St\'ephane Labrosse and Adrien Morison and Paul James Tackley},
     title = {Solid-state mantle convection coupled with a crystallising basal magma ocean},
     journal = {Comptes Rendus. G\'eoscience},
     pages = {5--21},
     publisher = {Acad\'emie des sciences, Paris},
     volume = {356},
     number = {S1},
     year = {2024},
     doi = {10.5802/crgeos.275},
     language = {en},
}

@ARTICLE{Morison2019,
       author = {{Morison}, A. and {Labrosse}, S. and {Deguen}, R. and {Alboussi{\`e}re}, T.},
        title = "{Timescale of overturn in a magma ocean cumulate}",
      journal = {Earth and Planetary Science Letters},
         year = 2019,
        month = jun,
       volume = {516},
        pages = {25-36},
          doi = {10.1016/j.epsl.2019.03.037},
       adsurl = {https://ui.adsabs.harvard.edu/abs/2019E&PSL.516...25M}
}

@ARTICLE{RobertsZhong2006,
       author = {{Roberts}, James H. and {Zhong}, Shijie},
        title = "{Degree-1 convection in the Martian mantle and the origin of the hemispheric dichotomy}",
      journal = {Journal of Geophysical Research (Planets)},
         year = 2006,
        month = jun,
       volume = {111},
       number = {E6},
          eid = {E06013},
        pages = {E06013},
          doi = {10.1029/2005JE002668},
       adsurl = {https://ui.adsabs.harvard.edu/abs/2006JGRE..111.6013R}
}

@ARTICLE{FischerGoddeKleine2017,
       author = {{Fischer-G{\"o}dde}, Mario and {Kleine}, Thorsten},
        title = "{Ruthenium isotopic evidence for an inner Solar System origin of the late veneer}",
      journal = {Nature},
         year = 2017,
        month = jan,
       volume = {541},
       number = {7638},
        pages = {525-527},
          doi = {10.1038/nature21045},
       adsurl = {https://ui.adsabs.harvard.edu/abs/2017Natur.541..525F}
}

@ARTICLE{Joiret2024,
       author = {{Joiret}, Sarah and {Raymond}, Sean N. and {Avice}, Guillaume and {Clement}, Matthew S.},
        title = "{Crash Chronicles: Relative contribution from comets and carbonaceous asteroids to Earth's volatile budget in the context of an Early Instability}",
      journal = {Icarus},
         year = 2024,
        month = may,
       volume = {414},
          eid = {116032},
        pages = {116032},
          doi = {10.1016/j.icarus.2024.116032},
archivePrefix = {arXiv},
       eprint = {2403.08545},
 primaryClass = {astro-ph.EP},
       adsurl = {https://ui.adsabs.harvard.edu/abs/2024Icar..41416032J}
}

@ARTICLE{Saurety2025,
       author = {{Saurety}, Adrien and {Caracas}, Razvan and {Raymond}, Sean N.},
        title = "{Impact-induced Vaporization during Accretion of Planetary Bodies}",
      journal = {The Astrophysical Journal Letters},
         year = 2025,
        month = mar,
       volume = {981},
       number = {1},
          eid = {L13},
        pages = {L13},
          doi = {10.3847/2041-8213/adb30e},
archivePrefix = {arXiv},
       eprint = {2502.04787},
 primaryClass = {astro-ph.EP},
       adsurl = {https://ui.adsabs.harvard.edu/abs/2025ApJ...981L..13S}
}

@ARTICLE{JohnsonMelosh2012,
       author = {{Johnson}, B.~C. and {Melosh}, H.~J.},
        title = "{Formation of spherules in impact produced vapor plumes}",
      journal = {Icarus},
         year = 2012,
        month = jan,
       volume = {217},
       number = {1},
        pages = {416-430},
          doi = {10.1016/j.icarus.2011.11.020},
       adsurl = {https://ui.adsabs.harvard.edu/abs/2012Icar..217..416J}
}

@ARTICLE{Martins2023,
       author = {{Martins}, Rayssa and {Kuthning}, Sven and {Coles}, Barry J. and {Kreissig}, Katharina and {Rehk{\"a}mper}, Mark},
        title = "{Nucleosynthetic isotope anomalies of zinc in meteorites constrain the origin of Earth{\textquoteright}s volatiles}",
      journal = {Science},
         year = 2023,
        month = jan,
       volume = {379},
       number = {6630},
        pages = {369-372},
          doi = {10.1126/science.abn1021},
       adsurl = {https://ui.adsabs.harvard.edu/abs/2023Sci...379..369M}
}

@ARTICLE{Martins2024,
       author = {{Martins}, Rayssa and {Morton}, Elin M. and {Kuthning}, Sven and {Goes}, Saskia and {Williams}, Helen M. and {Rehk{\"a}mper}, Mark},
        title = "{Primitive asteroids as a major source of terrestrial volatiles}",
      journal = {Science Advances},
         year = 2024,
        month = oct,
       volume = {10},
       number = {41},
          eid = {eado4121},
        pages = {eado4121},
          doi = {10.1126/sciadv.ado4121},
       adsurl = {https://ui.adsabs.harvard.edu/abs/2024SciA...10O4121M}
}

@ARTICLE{Burkhardt2021,
       author = {{Burkhardt}, Christoph and {Spitzer}, Fridolin and {Morbidelli}, Alessandro and {Budde}, Gerrit and {Render}, Jan H. and {Kruijer}, Thomas S. and {Kleine}, Thorsten},
        title = "{Terrestrial planet formation from lost inner solar system material}",
      journal = {Science Advances},
         year = 2021,
        month = dec,
       volume = {7},
       number = {52},
          eid = {eabj7601},
        pages = {eabj7601},
          doi = {10.1126/sciadv.abj7601},
archivePrefix = {arXiv},
       eprint = {2201.08092},
 primaryClass = {astro-ph.EP},
       adsurl = {https://ui.adsabs.harvard.edu/abs/2021SciA....7.7601B}
}

@ARTICLE{Goderis2021,
       author = {{Goderis}, Steven and {Sato}, Honami and {Ferri{\`e}re}, Ludovic and {Schmitz}, Birger and {Burney}, David and {Kaskes}, Pim and {Vellekoop}, Johan and {Wittmann}, Axel and {Schulz}, Toni and {Chernonozhkin}, Stepan M. and {Claeys}, Philippe and {de Graaff}, Sietze J. and {D{\'e}hais}, Thomas and {de Winter}, Niels J. and {Elfman}, Mikael and {Feignon}, Jean-Guillaume and {Ishikawa}, Akira and {Koeberl}, Christian and {Kristiansson}, Per and {Neal}, Clive R. and {Owens}, Jeremy D. and {Schmieder}, Martin and {Sinnesael}, Matthias and {Vanhaecke}, Frank and {Van Malderen}, Stijn J.~M. and {Bralower}, Timothy J. and {Gulick}, Sean P.~S. and {Kring}, David A. and {Lowery}, Christopher M. and {Morgan}, Joanna V. and {Smit}, Jan and {Whalen}, Michael T.},
        title = "{Globally distributed iridium layer preserved within the Chicxulub impact structure}",
      journal = {Science Advances},
         year = 2021,
        month = feb,
       volume = {7},
       number = {9},
        pages = {eabe3647},
          doi = {10.1126/sciadv.abe3647},
       adsurl = {https://ui.adsabs.harvard.edu/abs/2021SciA....7.3647G}
}

@ARTICLE{RosasKorenaga2018,
       author = {{Rosas}, Juan Carlos and {Korenaga}, Jun},
        title = "{Rapid crustal growth and efficient crustal recycling in the early Earth: Implications for Hadean and Archean geodynamics}",
      journal = {Earth and Planetary Science Letters},
         year = 2018,
        month = jul,
       volume = {494},
        pages = {42-49},
          doi = {10.1016/j.epsl.2018.04.051},
       adsurl = {https://ui.adsabs.harvard.edu/abs/2018E&PSL.494...42R}
}

@ARTICLE{Kimura1974,
       author = {{Kimura}, Kan and {Lewis}, Roy S. and {Anders}, Edward},
        title = "{Distribution of gold and rhenium between nickel-iron and silicate melts: implications for the abundance of siderophile elements on the Earth and Moon}",
      journal = {Geochimica et Cosmochimica Acta},
         year = 1974,
        month = may,
       volume = {38},
       number = {5},
        pages = {683-701},
          doi = {10.1016/0016-7037(74)90144-6},
       adsurl = {https://ui.adsabs.harvard.edu/abs/1974GeCoA..38..683K}
}

@ARTICLE{ReeseSolomatov2006,
       author = {{Reese}, C.~C. and {Solomatov}, V.~S.},
        title = "{Fluid dynamics of local martian magma oceans}",
      journal = {Icarus},
         year = 2006,
        month = sep,
       volume = {184},
       number = {1},
        pages = {102-120},
          doi = {10.1016/j.icarus.2006.04.008},
       adsurl = {https://ui.adsabs.harvard.edu/abs/2006Icar..184..102R}
}

@software{Anslow2026_zenodo,
  author       = {Anslow, Richard and
                  Landeau, Maylis and
                  Bonsor, Amy and
                  Itcovitz, Jonathan and
                  Shorttle, Oliver},
  title        = {The efficient delivery of highly-siderophile
                   elements to the core creates a mass accretion
                   catastrophe for the Earth
                  },
  month        = mar,
  year         = 2026,
  publisher    = {Zenodo},
  doi          = {10.5281/zenodo.18963669},
  url          = {https://doi.org/10.5281/zenodo.18963669},
}

%
%
%
%
%

\end{document}